\documentclass[aps,pra,twocolumn,superscriptaddress]{revtex4-2}
\usepackage{epsfig,amssymb, amsmath,fancyhdr,upgreek,color}
\newcommand{\bzeta}{{\boldsymbol \zeta}}
\newcommand{\bxi}{{\boldsymbol \xi}}
\newcommand{\bhata}{{\hat{\boldsymbol a}}} 

\begin{document}
\title{High-fidelity, quasi-deterministic entanglement generation using phase-matched spectral islands in a zero-added-loss multiplexing architecture}
\author{Jeffrey H. Shapiro}
\email{jhs@mit.edu}
\affiliation{Research Laboratory of Electronics, Massachusetts Institute of Technology, Cambridge, Massachusetts 02139 USA}
\author{Clark Embleton}
\email{cemblet2@uoregon.edu}
\affiliation{Department of Physics, Oregon Center for Optical, Molecular, and Quantum Science, University of Oregon, Eugene, Oregon 97403 USA}
\author{Michael G. Raymer}
\email{raymer@uoregon.edu}
\affiliation{Department of Physics, Oregon Center for Optical, Molecular, and Quantum Science, University of Oregon, Eugene, Oregon 97403 USA}
\author{Brian J. Smith}
\email{bjsmith@uoregon.edu}
\affiliation{Department of Physics, Oregon Center for Optical, Molecular, and Quantum Science, University of Oregon, Eugene, Oregon 97403 USA}
\date{October 9, 2025}

\begin{abstract}
Spontaneous parametric down-converters (SPDCs) are the best available entanglement sources for distributing entanglement in a quantum internet.  However, their intrinsically probabilistic nature, and their need to operate at low brightness to suppress multipair events, dictate that multiplexed SPDC arrays are required for high-rate distribution in that application.  Early SPDC multiplexing proposals involved path switching, whose switching losses significantly degrade performance.  The present paper proposes and analyzes a scheme for spectral multiplexing that provides entanglement-distribution rates well in excess of the state of the art. It builds on zero-added-loss multiplexing (ZALM)~[Phys. Rev. Appl. {\bf 19}, 054029 (2023)] for high-rate heralded entanglement generation, which does not require a switched array of SPDCs.    Our ZALM's SPDCs rely on nonlinear crystals with $N_I$ phase-matched spectral islands, each generating two-mode squeezed-vacuum states. Also, our ZALM's multiplexing protocol uses both same-island and cross-island heralding, which allows the entanglement-delivery rate to approximately scale as $N_I^2$ in the realistic weak-squeezing regime.  As a result, our scheme uses an order of magnitude fewer spectral channels than the original ZALM proposal, which may enable near-term implementations of satellite-to-ground or fiber-optic based ZALM architectures.
\end{abstract}

\maketitle

\section{Introduction \label{Intro}}
Qubit teleportation~\cite{Bennett1993,Bouwmeester1997,Riebe2004} is a key primitive for quantum communication because it affords error-free, repeaterless transmission of quantum states over pure-loss bosonic channels with $\eta< 1/2$ transmissivities, viz., links with zero quantum capacity~\cite{Weedbrook2012}.  At power levels below the onset of nonlinear effects, and employing post-propagation dispersion compensation, $L$-km-long transmission over standard single-mode fiber at 1.55\,$\upmu$m wavelength may be regarded as a pure-loss channel~\cite{footnote0} with transmissivity $\eta = 10^{-0.02L}$.  Hence its quantum capacity vanishes for $L > 15\,$km.  Because the fiber channel's classical capacity is not subject to this catastrophic collapse, the internet's fiber backbone uses erbium-doped fiber amplifiers (EDFAs)---spaced 80--100\,km apart---in lieu of classical repeaters.  Thus, because co-locating quantum repeaters with those amplifiers is highly desirable for efficient build-out~\cite{Toudeh-Fallah2024} of a quantum internet~\cite{Lloyd2004,Kimble2008,Wehner2018}, transmission via qubit teleportation becomes necessary, and that requires creating reservoirs of shared qubit entanglement between these EDFA sites.  

Spontaneous parametric down-converters (SPDCs) have long been workhorses for  generating polarization-entangled~\cite{Kwiat1995,Kwiat1999,Kim2001,Kurtsiefer2001} and time-bin entangled~\cite{Brendel1999,Marcikic2002,Marcikic2004} photon pairs.  Polarization entanglement is preferable
for satellite-to-ground entanglement distribution, because
atmospheric turbulence is not depolarizing~\cite{Andrews1998}. For terrestrial entanglement distribution over optical
fiber, however, time-bin entanglement is the right choice, because standard fiber does not preserve polarization and polarization-maintaining fiber is currently too lossy
for long-distance operation.  Sadly, in both cases, record-setting demonstrations leave much to be desired with respect to entanglement distribution over long distances.  For example, Neumann~\emph{et al}.~\cite{Neumann2022} realized only a 9\,s$^{-1}$ distribution rate over 248\,km of deployed fiber, while Yin~\emph{et al}.~\cite{Yin2017} achieved only a 1.1\,s$^{-1}$ distribution rate from the Micius satellite to a pair of ground stations separated by 1200\,km.  

The preceding demonstrations  used bidirectionally-pumped Sagnac SPDCs~\cite{Kim2006,Wong2006}, which are renowned for their excellent stability and high efficiency.  Nevertheless, a Sagnac SPDC---like SPDCs in general---is  intrinsically probabilistic, a characteristic that adversely impacts its entanglement-distribution rate.  In particular, to suppress unwanted multipair events, SPDCs are typically operated at low brightness, e.g., an average of 0.01 entangled pairs per pump pulse.  Multiplexing an array of SPDCs could push up the entangled-pair generation rate to near-deterministic, or at least quasi-deterministic, performance.  But multiplexing schemes that rely on path switching at the source, such as Mower~\emph{et al}.~\cite{Mower2011} and Dhara~\emph{et al}.~\cite{Dhara2022}, suffer significant performance penalties from their switch losses.  Chen~\emph{et al}.'s zero-added-loss multiplexing (ZALM)~\cite{Chen2023} is different.  ZALM uses the equivalent of two Sagnac sources~\cite{footnote1} and partial Bell-state measurements (BSMs)~\cite{Braunstein1995} to herald the generation of polarization-entangled biphotons across
a large number of frequency-multiplexed signal channels
that collectively span the down-converters' multi-THz phase-matching bandwidth~\cite{footnote2}. According to Chen~\emph{et al}., this arrangement has the potential to keep multipair emissions negligible on a per-channel basis while approaching quasi-deterministic performance over the SPDCs' full phase-matching bandwidth.  

ZALM uses a source-in-the-middle configuration, which has higher efficiency than its meet-in-the-middle and sender-receiver alternatives~\cite{Jones2016}.  For the quantum internet's fiber backbone, every classical-internet EDFA site would contain a ZALM quantum transmitter (QTX) to distribute entangled photon pairs to ZALM quantum receivers (QRXs) at that site's nearest neighbors, Alice and Bob.   As in Thomas~\emph{et al}.~\cite{Thomas2024}, classical traffic---including the identities of the frequency channels containing the heralded biphotons---could be confined to the C-band, while optical transmission of the heralded biphotons could occur in the O-band.    Alice and Bob's QRXs would then employ quantum mode conversion---frequency conversion~\cite{Kumar1990,Huang1992} and bandwidth compression~\cite{Karpinski2017}---to efficiently couple the biphoton received in each heralded channel to intracavity color-center quantum memories.  ZALM thus avoids path-switching losses in its QTX---hence justifying its ``zero-added-loss'' name---while relying on high-efficiency quantum-state frequency conversion~\cite{Wang2018} at Alice and Bob's QRXs to match the incoming light to their quantum memories' input frequency. 
 
Chen~\emph{et al}.~\cite{Chen2023} presented a broad treatment of ZALM, including its source, mode-conversion, and memory-loading characteristics, as well as its overall performance scaling.  Their ZALM relied on a burst of pump pulses to generate a striped frequency-domain biphoton wave function over a very broad (10\,THz) bandwidth that, together with narrowband (1\,GHz) filtering in the partial BSM and suitable (12.5\,GHz) channelization at the receivers, led to a large number (800) of channels whose frequency-domain biphoton wave functions each approximated that of a spectrally-factorable, i.e., single-temporal-mode, state.  Such near single-temporal-mode biphoton generation ensured the production of high-fidelity ($\ge$\,99\%) biphotons, and the 800 channels provided quasi-deterministic ($\ge$\,25\% generation probability per pump burst) operation.  Unfortunately, Chen~\emph{et al}.'s analysis did not fully account for the combined effects of multipair events and losses.   

To overcome these limitations, we take the essence of Chen~\emph{et al}.'s approximately-factorable  channelized biphoton wave function to its ideal limit.  Motivated by Morrison~\emph{et al}.~\cite{Morrison2022}, who domain engineered a $\chi^{(2)}$ crystal, pumped by a transform-limited Gaussian pulse with an appropriately chosen bandwidth, to realize a biphoton wave function with 8 discrete and spectrally factorable frequency bins, we assume the ZALM source's SPDCs each have a modest number ($\ll$\,800) of phase-matched spectral islands that each generate two-mode squeezed-vacuum states.  As a result,  our analysis is able to go well beyond Chen~\emph{et al}.'s, in that ours includes multipair events of all orders, losses in the partial BSM and in propagation to Alice and Bob's QRXs, and the use of both same-island and cross-island heralding.  More importantly, it shows that entangled photon pairs can be generated and delivered with high ($\ge$\,99\%) Bell-state fidelity and very high ($\ge$\,99.9\%) Bell-state fraction despite losses in the partial BSM and in propagation.  Furthermore, the generation rate can be quasi-deterministic ($\ge$\,25\% generation probability per pump pulse) with a modest number ($\ll$\,800) of islands.  

The remainder of the paper is organized as follows.  Section~\ref{islands} introduces the islands-based ZALM source and shows that in ideal, i.e., lossless, operation it can easily generate entanglement that has high fidelity and is near-deterministic.  Section~\ref{lossy_operation} extends our treatment of islands-based ZALM to include losses in the partial BSM and in the signal photons' propagation to Alice and Bob's QRXs.  Here, same-island plus cross-island heralding permits the QTX's entanglement generation to be quasi-deterministic, while delivering entangled pairs to Alice and Bob's QRXs with high Bell-state fidelity and very high Bell-state fraction even with 1\% QTX-to-QRX transmissivity.  Concluding remarks are presented in Sec.~\ref{discussion}, along with a glossary of principal symbols.  Derivations underlying the performance metrics reported in Secs.~\ref{lossy_BSM} and \ref{lossy_prop} appear in Appendices~\ref{AppendA} and \ref{AppendB}, respectively.

\section{Lossless ZALM with phase-matched spectral islands \label{islands}}

\begin{figure}[hbt]
    \centering
\includegraphics[width=0.49\textwidth]{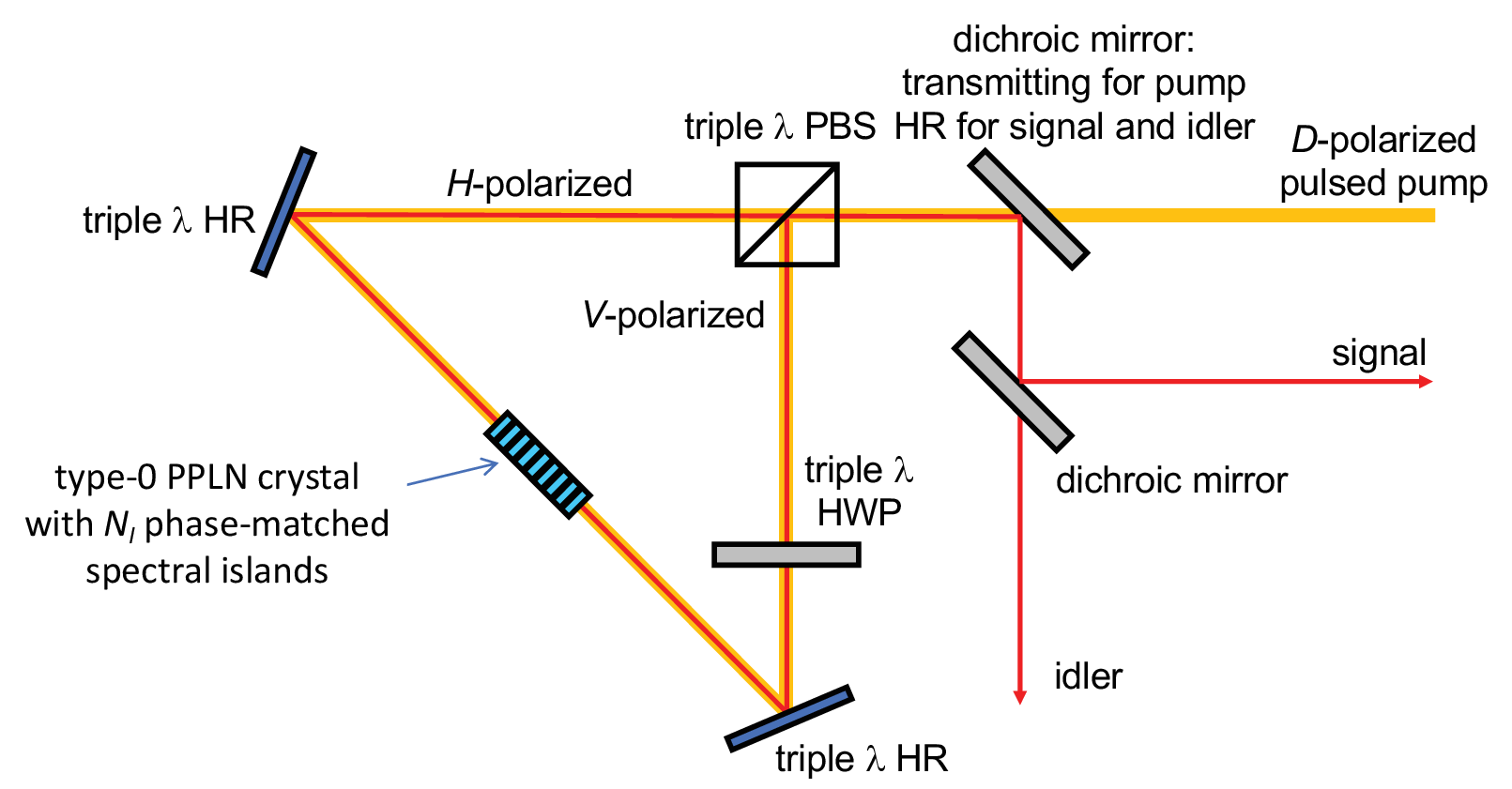}
    \caption{Schematic of a Sagnac-configured SPDC source~\cite{Wong2006} of signal-idler biphotons suitable for use in islands-based ZALM.  A periodically-poled lithium niobate (PPLN) crystal with $N_I$ phase-matched spectral islands~\cite{footnote3} is bidirectionally pulse-pumped for type-0 nondegenerate phase matching.  $D$, $H$, and $V$: diagonal, horizontal, and vertical polarizations.  HR:  high reflector. $\lambda$:  wavelength.  PBS:  polarizing beam splitter.  HWP:  half-wave plate. \label{Sagnac-source_fig}}    
\end{figure}

\begin{figure}[hbt]
    \centering
\includegraphics[width=0.99\columnwidth]{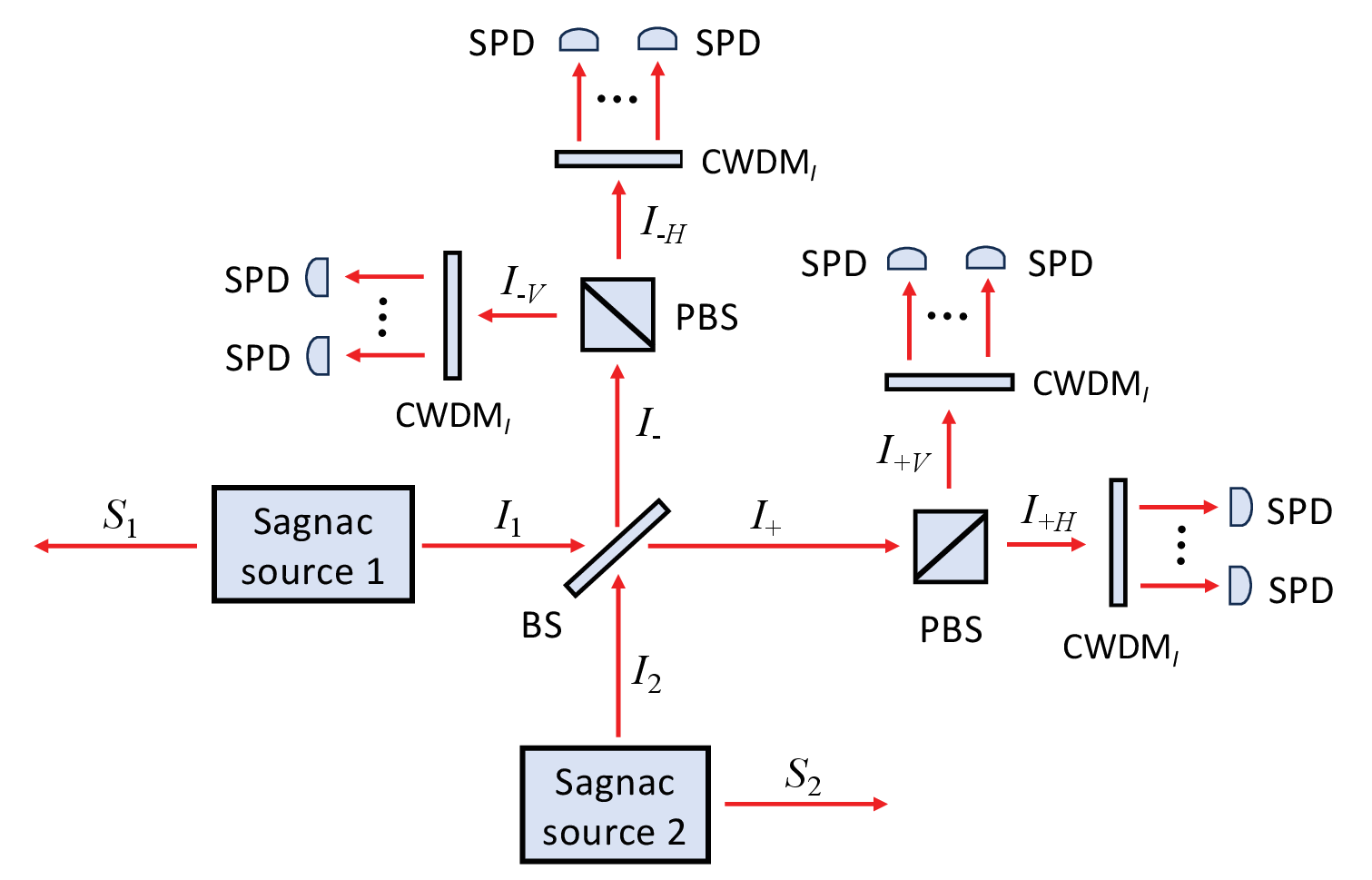}
    \caption{Schematic of islands-based ZALM's partial Bell-state measurement for heralding polarization-entangled photon pairs. Here $S_k$ and $I_k$ for $k = 1,2$ denote the signal ($S$) and idler ($I$) beams from the $k$th Sagnac source.  $I_\pm$ denote the idler-beam outputs from the 50--50 beam splitter (BS); $I_{\pm P}$ for $P= H, V$ denote the horizontally ($H$) and vertically ($V$) polarized outputs from the polarizing beam splitter (PBS) illuminated by $I_\pm$;  CWDM$_I$ denotes the  idler-beam coarse wavelength-division multiplexing filter.  SPD,  single-photon detector.
 \label{partial-BSM_fig} }
\end{figure}

To set the stage, we first introduce and analyze islands-based ZALM in the ideal case of lossless operation.  Modifying figures  from our previous treatment of dense wavelength-division multiplexed (DWDM) ZALM using type-0 phase matching~\cite{Shapiro2024}, islands-based ZALM's QTX is shown schematically in Figs.~\ref{Sagnac-source_fig} and \ref{partial-BSM_fig} and explained below.

\begin{figure}[hbt]
    \centering
\includegraphics[width=0.5\columnwidth]{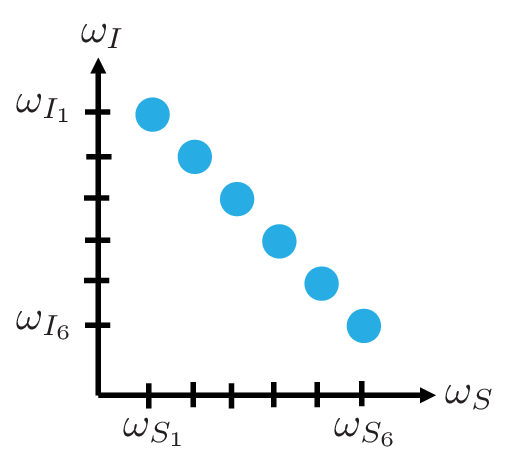}
    \caption{Sketch of the frequency-domain wave function for a biphoton produced by 6 identical, spectrally-factorable phase-matched spectral islands, with signal-idler center frequencies $\{(\omega_{S_n},\omega_{I_n}): n=1,2,\ldots,6\}$.
 \label{islands_sketch} }
\end{figure}

Islands-based ZALM's QTX uses a pair of Sagnac sources that employ identical periodically-poled lithium niobate (PPLN) crystals each with $N_I$ phase-matched spectral islands.  The SPDCs' islands are type-0 phase matched and pulse-pumped in phase to produce signal-idler pairs from each island that, except for their island-dependent signal and idler center frequencies, are in identical temporal modes. To achieve such a state, the spectral islands must be identical, nonoverlapping, and spectrally factorable, i.e., expressible as a product of functions of the signal frequency and the idler frequency~\cite{Grice1997} as sketched in Fig.~\ref{islands_sketch} for 6 islands.

Restricting our attention to the foregoing temporal modes, and assuming operation in the SPDCs' usual no-pump-depletion regime, we have that the output states from Sagnacs  $k=1,2$ resulting from a single pump pulse are the following tensor products of two-mode squeezed-vacuum (TMSV) states,
\begin{align}
|\psi\rangle_{S_kI_k} = &\bigotimes_{n=1}^{N_I}\sum_{m=0}^\infty\sqrt{P_m}\,|m\rangle_{S_{k_{n_H}}}|m\rangle_{I_{k_{n_H}}} \nonumber\\[.05in]
&\bigotimes_{n'= 1}^{N_I}\sum_{m'=0}^\infty\sqrt{P_{m\prime}}\,|m'\rangle_{S_{k_{n'_V}}}|m'\rangle_{I_{k_{n'_V}}}.
\label{islands_state}
\end{align}
Here: $|m\rangle_{K_{n_P}}$ is the $m$-photon Fock state of the $P$-polarized ($P=H,V$) signal ($K=S$) and idler ($K=I$) modes emitted by the $n$th phase-matching island; and 
\begin{equation}
P_m \equiv \frac{(G-1)^m}{G^{m+1}},\mbox{ for $m=0,1,\ldots, \infty$},
\label{BoseEinstein}
\end{equation} 
is the Bose-Einstein probability distribution with $G$ being the squeezing gain, so that $G-1 > 0$ is the average number of signal-idler pairs emitted per SPDC-island per pump pulse.  The joint signal-idler state of the two Sagnacs is then $|\psi\rangle_{{\bf SI}} = |\psi\rangle_{S_1I_1}|\psi\rangle_{S_2I_2}.$ 

Figure~\ref{partial-BSM_fig}'s partial-BSM setup first combines the idlers from the two Sagnacs on a 50--50 beam splitter, after which it separates their horizontal ($H$) and vertical ($V$) polarization components with polarizing beam splitters (PBSs).  Coarse wavelength-division multiplexing (CWDM) filters then perfectly separate the modes from the different phase-matching islands, and the CWDM filters' outputs illuminate single-photon detectors (SPDs) that have partial number-resolution (PNR), i.e., 0, 1, or $>1$, capability and no dark counts.  A same-island herald from a particular pump pulse is declared for the $n$th island when exactly two idler photons are detected from the $n$th island, with one being $H$ polarized and the other $V$ polarized.  If those detections both occur in Fig.~\ref{partial-BSM_fig}'s $I_+$ branch or in its $I_-$ branch, the signal modes' heralded state is the polarization Bell-state triplet~\cite{Shapiro2024}
\begin{equation}
|\psi^+\rangle_{S_{1_n}S_{2_n}} = (|1\rangle_{S_{1_{n_H}}}|1\rangle_{S_{2_{n_V}}}+ |1\rangle_{S_{1_{n_V}}}|1\rangle_{S_{2_{n_VH}}})/\sqrt{2}.
\end{equation}
Alternatively, if one of those detections occurs in Fig.~\ref{partial-BSM_fig}'s $I_+$ branch, while the other occurs in its $I_-$ branch, the signal modes' heralded state is the polarization Bell-state singlet
\begin{equation}
|\psi^-\rangle_{S_{1_n}S_{2_n}} = (|1\rangle_{S_{1_{n_H}}}|1\rangle_{S_{2_{n_V}}}-|1\rangle_{S_{1_{n_V}}}|1\rangle_{S_{2_{n_VH}}})/\sqrt{2}.
\end{equation}

Because here we are assuming lossless optics and unit-efficiency, PNR-capable detectors with no dark counts in our partial-BSM apparatus, it is clear that we will get an $n$th-island herald from a particular pump pulse if and only if two $n$th-island idler photons enter the partial-BSM apparatus with one being $H$-polarized and the other being $V$-polarized.  There are four ways---all equally likely---in which this $n$th-island heralding event, $\mathcal{H}_n$, can occur:  Sagnacs~1 and 2 emit, respectively, an $n$th-island $H$-polarized idler photon and an $n$th-island $V$-polarized idler photon, or vice versa; and Sagnac~1 emits an $n$th-island $H$-polarized idler photon and an $n$th-island $V$-polarized idler photon with no $n$th-island emission from Sagnac~2, or vice versa.  The probability of an $n$th-island herald from a particular pump pulse 
is therefore~\cite{footnoteA}
\begin{equation}
\Pr(\mathcal{H}_n) = 4(G-1)^2/G^6
\label{lossless_herald}
\end{equation}
and independent of $n$ because we have assumed all modes have the same squeezing gain, cf.\@ Eq.~(\ref{islands_state}).
Note that only the first two $n$th-island heralding possibilities are \emph{true} heralds, i.e., heralds for which the heralded $|\psi^\pm\rangle_{S_{1_n}S_{2_n}}$ state is indeed the one that is transmitted to Alice and Bob.  The second two $n$th-island heralding possibilities are called \emph{false} heralds, viz., they result in two $n$th-island signal photons being sent to Alice and none to Bob, or vice versa.  Both the true and false $n$th-island heralds arise because the partial-BSM's 50--50 beam splitter erases which-path information from the ZALM QTX's two Sagnacs.  

At this point we introduce the three metrics by which we will measure the ZALM QTX's performance.  The first is the per-pump-pulse probability that Alice and Bob are sent a true $n$th-island herald, $\Pr(\mathcal{H}_{\rm true})$, which, like $\Pr(\mathcal{H}_n)$, is independent of $n$.  The second metric is the Bell-state fraction, $\mathcal{B}$,  of the state transmitted after a true $n$th-island herald, i.e., the conditional probability that this state lies in the $S_{1_n}S_{2_n}$ Hilbert space spanned by those modes' polarization Bell states given that the QTX's output state is \emph{loadable}, viz., it sends photons to both Alice and Bob.  The third metric is the Bell-state fidelity, $\mathcal{F}$, which is the conditional probability that the state sent to Alice and Bob is the state that was heralded, given a true $n$th-island herald has occurred and the state sent to Alice and Bob lies in the $S_{1_n}S_{2_n}$ polarization Bell-state Hilbert space.
 
\begin{figure}[hbt]
    \centering
\includegraphics[width=0.99\columnwidth]{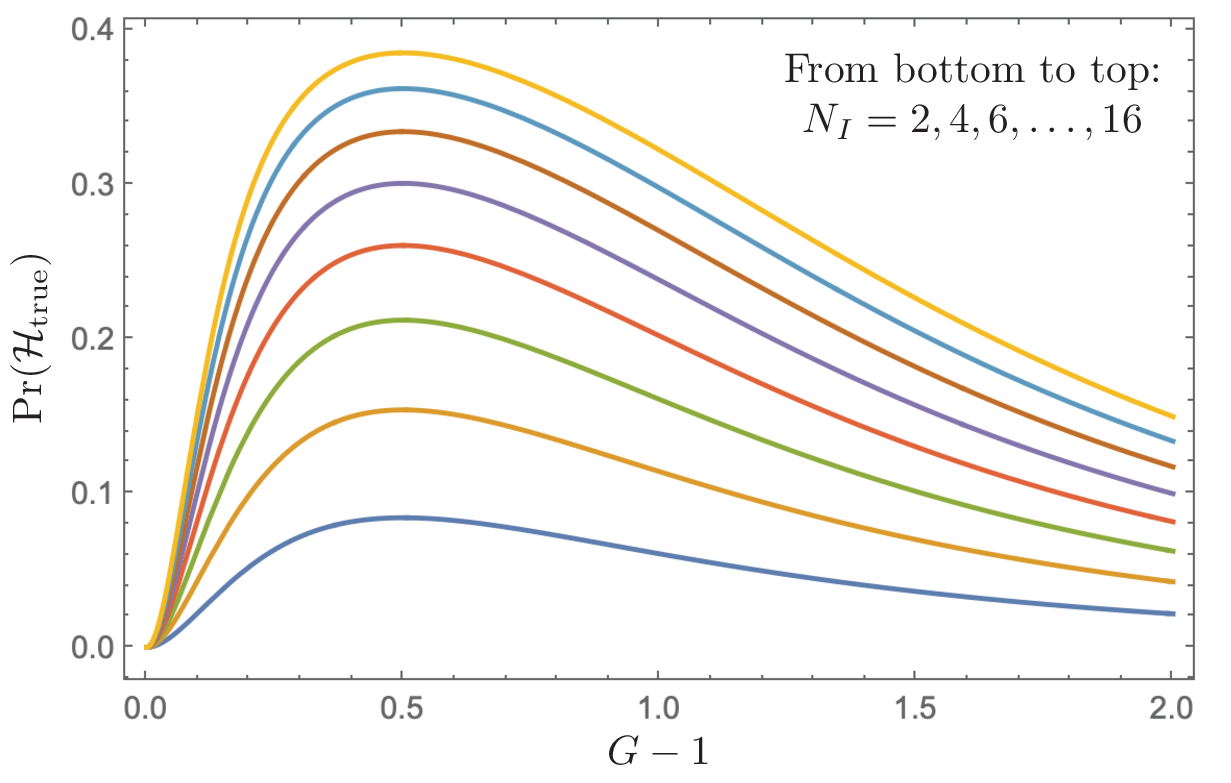}
    \caption{Performance of ideal islands-based ZALM with same-island heralding: per-pump-pulse probability of an $n$th-island true herald, $\Pr(\mathcal{H}_{\rm true})$, versus average number of signal-idler pairs per SPDC island per pump pulse, $G-1$, for $N_I = 2,4,6,\ldots, 16$ islands. \label{Lossless_Ptrue} }
\end{figure}

To evaluate these metrics for same-island heralding we will assume the ZALM QTX sends at most one photon-pair herald per pump pulse to Alice and Bob, in keeping with a ZALM QRX architecture that allocates only one qubit of quantum memory to each pump pulse. The probability that Alice and Bob are sent a true $n$th-island herald---and hence a polarization-entangled photon pair---on a particular pump pulse in this case is therefore~\cite{footnoteB}
\begin{equation}
\Pr(\mathcal{H}_{\rm true}) = \{1 - [1-\Pr(\mathcal{H}_n)]^{N_I}\}/2,
\label{true-herald}
\end{equation}
i.e., half the probability that at least one island heralds.  
Our assumption of lossless QTX optics and unit-efficiency, PNR-capable SPDs with no dark counts immediately implies, from Eq.~(\ref{islands_state}), that a $|\psi^\pm\rangle_{S_{1_n}S_{2_n}}$ true herald results in a $|\psi^\pm\rangle_{S_{1_n}S_{2_n}}$ biphoton being transmitted.  So, the ideal islands-based ZALM QTX with same-island heralding realizes unit Bell-state fraction ($\mathcal{B} = 1$) and unit Bell-state fidelity ($\mathcal{F}  = 1$).  Figure~\ref{Lossless_Ptrue} plots the per-pump-pulse probability of a true herald versus the average number of signal-idler pairs per SPDC island per pump pulse for $N_I = 2, 4, 6, \ldots, 12$ islands.  

The preceding results tell us that the ideal ZALM QTX with same-island heralding and at most one herald per pump pulse is near perfect, i.e., it has unit Bell-state fraction and unit Bell-state fidelity with a true-herald per-pump-pulse probability that exceeds 0.25 for 8 islands at $G-1 = 0.5$, and is more than 80\% of that metric's ultimate (0.5) limit for 16 islands at $G-1= 0.5$.  In contrast, the lossless ZALM QTX from Chen~\emph{et al}.~\cite{Chen2023} needed 800 channels to achieve high-fidelity operation with 0.25 per-pump-pulse probability of a true herald.  

\section{Lossy ZALM with phase-matched spectral islands \label{lossy_operation}}
Having painted such a rosy picture for islands-based ZALM QTX with ideal equipment, same-island heralding, and at most one herald per pump pulse, it behooves us to explore how rapidly that QTX's performance degrades in non-ideal operation.  We should also quantify the quality of the state delivered to Alice and Bob after loss in propagation from the ZALM QTX to their QRXs.  Thus, in this section, we first consider a lossy QTX whose losses are symmetrically distributed on the idler paths through the partial-BSM setup and lumped together into sub-unit quantum efficiencies, $\eta_T$, at each SPD~\cite{footnote4}, but with no losses in the Sagnac sources' signal paths.  The second case we treat augments the preceding partial-BSM losses with symmetrically-distributed sub-unit signal-path transmissivities, $\eta_R$, from the ZALM QTX to Alice and Bob's QRXs.  

The per-pump-pulse $n$th-island heralding probabilities for both cases described above coincide, because they  depend only on the reduced density operator for the four idler modes entering the partial BSM and the SPDs' quantum efficiency.  From Eq.~(\ref{islands_state}), the $4N_I$  signal-idler mode pairs, $\{(S_{k_{n_P}},I_{k_{n_P})}\}$ are in independent, identically-distributed (iid) two-mode squeezed-vacuum states.  Tracing out the signal states leaves the $\{I_{k_{n_P}}\}$ modes in iid thermal states with average photon number $G-1$.  Passing those idler modes through Fig.~\ref{partial-BSM_fig}'s 50--50 beam splitter produces idler modes, $\{I_{\pm P_{n}}\}$, that are also in iid thermal states with average photon number $G-1$.  The well-known result for thermal-state photodetection with sub-unit quantum efficiency then gives~\cite{footnoteC}
\begin{equation}
\Pr(\mathcal{H}_n) =  \frac{4[\eta_T(G-1)]^2}{[\eta_T(G-1)+1]^6},
\label{lossy_herald}
\end{equation}
for the per-pump-pulse probability that an $n$th-island $H$-polarized idler photon and an $n$th-island $V$-polarized photon are detected.  Moreover, as before, half of these detections are false heralds.  Furthermore, continuing to assume that the ZALM QTX transmits at most one herald per pump pulse, we have that the per-pump-pulse probability of an $n$th-island true herald is given by Eq.~(\ref{true-herald}) with $\Pr(\mathcal{H}_n)$ now from Eq.~(\ref{lossy_herald}) instead of Eq.~(\ref{lossless_herald}).  

There is no need to replot $\Pr(\mathcal{H}_{\rm true})$ for a lossy partial BSM; we need only change $G-1$ to $\eta_T(G-1)$ on Fig.~\ref{Lossless_Ptrue}'s abscissa.  That said, before using $G-1 = 1/(2 \eta_T)$ to get $\Pr(\mathcal{H}_{\rm true}) > 0.25$ for an 8-island QTX with a sub-unit $\eta_T$, we must evaluate that transmissivity's impact on the Bell-state fraction and Bell-state fidelity.  This need arises because multipair events are increasingly important as $G$ increases, and $\eta_T < 1$ implies that the SPDs do not completely ward them off.   Finding $\mathcal{B}$ and $\mathcal{F}$ for a sub-unit $\eta_T$ is the task of Sec.~\ref{lossy_BSM}.  

\subsection{Islands-based ZALM with partial-BSM loss \label{lossy_BSM}} 

Our path to finding the Bell-state fraction and Bell-state fidelity for a ZALM QTX with sub-unit detection efficiency runs through the transform domain, starting from the anti-normally-ordered characteristic function associated with $\hat{\rho}_{{\bf S}{\bf I}}$, the density operator for the signal, ${\bf S} = (S_1,S_2)$, and idler, ${\bf I} = (I_1,I_2)$, modes of a single island, 
\begin{equation}
\chi_A^{\rho_{\bf SI}}(\bzeta) \equiv 
{\rm Tr}\!\left[\hat{\rho}_{\bf SI}e^{-\bzeta^\dagger\bhata}e^{\bhata^\dagger\bzeta}\right],
\label{antinorm_start}
\end{equation} 
where, for convenience, we suppress the island index, $n$.
In Eq.~(\ref{antinorm_start}): $\bzeta^\dagger \equiv \left[\begin{array}{cc}\bzeta_S^\dagger & \bzeta_I^\dagger\end{array}\right]$ and $\bhata^\dagger \equiv \left[\begin{array}{cc}\bhata_S^\dagger & \bhata_I^\dagger\end{array}\right]$ with 
\begin{equation}
\bzeta^\dagger_K \equiv \left[\begin{array}{cccc}\zeta^*_{K_{1_H}} & \zeta^*_{K_{1_V}} & \zeta^*_{K_{2_H}} & \zeta^*_{K_{2_V}}\end{array}\right],\mbox{ for $K=S,I$}, 
\end{equation}
and
\begin{equation}
\bhata_K^\dagger \equiv \left[\begin{array}{cccc}\hat{a}^\dagger_{K_{1_H}} & \hat{a}^\dagger_{K_{1_V}} & \hat{a}^\dagger_{K_{2_H}} & \hat{a}^\dagger_{K_{2_V}} \end{array}\right], \mbox{ for $K=S,I$}
\end{equation}
being four-dimensional (4D) row vectors of complex-valued parameters and photon creation operators, respectively.  The density operator $\hat{\rho}_{\bf SI}$ can be recovered from its anti-normally-ordered characteristic function via the 16D operator-valued inverse Fourier transform~\cite{footnote5},
\begin{equation}
\hat{\rho}_{\bf SI} = \int\!\frac{{\rm d}^{16}\bzeta}{\pi^8}\,\chi_A^{\rho_{\bf SI}}(\bzeta)e^{-\bhata^\dagger\bzeta}e^{\bzeta^\dagger\bhata}.
\label{inverse_transform}
\end{equation}

A standard squeezed-state calculation, using the $n$th-island state from Eq.~(\ref{islands_state}), gives~\cite{Weedbrook2012}
\begin{align}
\chi&_A^{\rho_{\bf SI}}(\bzeta) = \nonumber \\[.05in]
&\hspace*{.1in} \exp\!\left[-G\bzeta^\dagger\bzeta + 2\sqrt{G(G-1)}\,{\rm Re}(\bzeta_S^T\bzeta_I)\right]. 
\label{chiAstart}
\end{align}
Appendix~\ref{AppendA} uses the mode transformation for 50--50 beam splitting and the mode transformation that accounts for sub-unit detector efficiency to obtain the anti-normally-ordered characteristic function associated with the conditional density operator for the ${\bf S}$ modes, $\hat{\rho}_{{\bf S}\mid I'_{+H}I'_{-V}}$, given that an $I'_{+H}I'_{-V}$ herald has occurred:
\begin{align}
\chi_A^{\rho_{{\bf S}\mid I'_{+H}I'_{-V}}}(\bzeta_S) &= e^{-\bzeta_S^\dagger \bzeta_S/N_S}\left[1 - \frac{|\zeta_{S_{1_H}}+\zeta_{S_{2_H}}|^2}{2N_S}\right]\nonumber \\[.05in] & \times \left[1 - \frac{|\zeta_{S_{1_V}}-\zeta_{S_{2_V}}|^2}{2N_S}\right].
\label{CondxCharFn}
\end{align}
Here: ${\bf I}'\equiv \{I'_{+H},I'_{+V},I'_{-H},I'_{-V}\}$ are the efficiency-$\eta_T$ detected idler modes of the single island under consideration; and $N_S \equiv [\eta_T(G-1)+1]/G.$  Using the operator-valued inverse Fourier transform of this characteristic function, Appendix~\ref{AppendA} goes on to show that  
\begin{equation}
{}_{\bf S}\langle \psi^-|\hat{\rho}_{{\bf S}\mid I'_{+H}I'_{-V}}|\psi^-\rangle_{\bf S} = N_S^6/2,
\label{s_nIV.a}
\end{equation}
and
\begin{align}
 {}_{\bf S}\langle \psi^+|\hat{\rho}_{{\bf S}\mid I'_{+H}I'_{-V}}&|\psi^+\rangle_{\bf S} = 
 {}_{\bf S}\langle \phi^+|\hat{\rho}_{{\bf S}\mid I'_{+H}I'_{-V}}|\phi^+\rangle_{\bf S} \nonumber \\[.05in]
 & = {}_{\bf S}\langle \phi^-|\hat{\rho}_{{\bf S}\mid I'_{+H}I'_{-V}}|\phi^-\rangle_{\bf S} 
 = 0,
 \label{e_nIV.a}
 \end{align}
are the polarization Bell-state probabilities implied by $\hat{\rho}_{{\bf S}\mid I'_{+H}I'_{-V}}$.
 
Taken together, Eqs.~(\ref{s_nIV.a}) and (\ref{e_nIV.a}) imply that, given there has been an $I'_{+H}I'_{-V}$ herald, the $S_1S_2$ state sent to Alice and Bob has unit Bell-state fidelity.  To see that this is so, remember that an $I'_{+H}I'_{-V}$ herald tells Alice and Bob to expect a $|\psi^-\rangle_{\bf S}$ state, and that the Bell-state fidelity for the $I'_{+H}I'_{-V}$ herald is defined to be
${}_{\bf S}\langle \psi^-|\hat{\rho}^{(\mathcal{H}_B)}_{{\bf S}\mid I'_{+H}I'_{-V}}|\psi^-\rangle_{\bf S}$,
where $\mathcal{H}_B$ is the Hilbert space of the $S_1S_2$ states spanned by their polarization Bell states and $\hat{\rho}^{(\mathcal{H}_B)}_{{\bf S}\mid I'_{+H}I'_{-V}}$ is the normalized conditional density operator for the $S_1S_2$ states in $\mathcal{H}_B$ given there has been an $I'_{+H}I'_{-V}$ herald. 
Thus, because Eqs.~(\ref{s_nIV.a}) and (\ref{e_nIV.a}) show that the $|\psi^-\rangle_{\bf S}$ polarization Bell-state singlet is the \emph{only} Bell state transmitted to Alice and Bob when the $I'_{+H}I'_{-V}$ herald occurs, unit Bell-state fidelity ensues.  Multipair events do \emph{not} affect the Bell-state fidelity of a lossy partial BSM because, at this point, we have assumed there are no losses in the signal paths.  Note too that, as outlined in Appendix~\ref{AppendA}, the other three heralding possibilities---$I'_{+V}I'_{-H}$ and $I'_{\pm H}I'_{\pm V}$---also result in $S_1S_2$ states with unit Bell-state fidelity.  Unlike the Bell-state fidelity, however, the Bell-state fraction of the ZALM QTX's output state \emph{is} adversely impacted by a sub-unit $\eta_T$. Moreover, quantifying that impact, as we will do next, is crucial in that unit Bell-state fidelity is worthless if the Bell-state fraction of the ZALM QTX's $S_1S_2$ outputs is minuscule, i.e., if those outputs are completely dominated by multipair events.  We shall see in Sec.~\ref{lossy_prop}, however, that the loss encountered in long-distance fiber propagation or satellite-to-ground transmission makes it possible to have both high Bell-state fidelity and exceedingly-high Bell-state fraction.

There are two classes of $S_1S_2$ output states that are outside the polarization Bell-states' Hilbert space:  output states due to false heralds, and output states due to multipair events in which signal photons are sent to both Alice and Bob.  Defining a \emph{memory-loadable} output state to be one that sends signal photons to both Alice and Bob, the partial-BSM's Bell-state fraction, given an $I'_{+H}I'_{-V}$ herald, is $\Pr({\rm Bell}\mid I'_{+H}I'_{-V})/\Pr({\rm loadable}\mid I'_{+H}I'_{-V})$.   
From Eqs.~(\ref{s_nIV.a}) and (\ref{e_nIV.a}) we have that
\begin{equation}
\Pr({\rm Bell}\mid I'_{+H}I'_{-V}) = N_S^6/2.
\end{equation}
In Appendix~\ref{AppendA} we show that
\begin{equation}
\Pr({\rm loadable}\mid I'_{+H}I'_{-V}) = 1- N_S^2/2.
\end{equation}
Moreover, these probabilities also apply to the other heralding possibilities, $I'_{-H}I'_{+V}$ and $I'_{\pm H}I'_{\pm V}$.  It follows that the lossy partial-BSM's Bell-state fraction is
\begin{equation}
\mathcal{B} = \frac{N_S^6}{2-N_S^2}.
\end{equation}

\begin{figure}[hbt]
    \centering
\includegraphics[width=0.99\columnwidth]{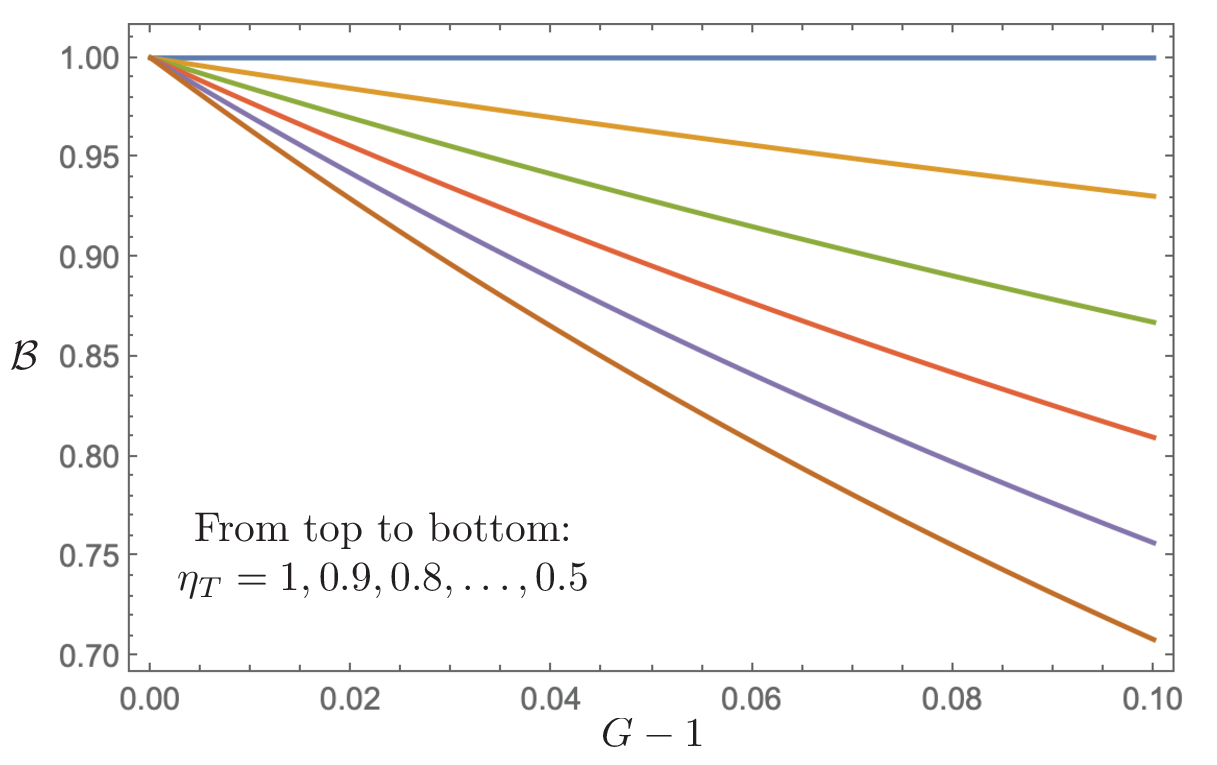}
    \caption{Performance of islands-based ZALM with a lossy partial BSM: Bell-state fraction, $\mathcal{B}$, of the $S_1S_2$ output state versus average number of signal-idler pairs per SPDC island per pump pulse, $G-1$, for (top to bottom) $\eta_T= 1,0.9,0.8,\ldots, 0.5$. \label{BellFractionT} }
\end{figure}

Figure~\ref{BellFractionT} plots the Bell-state fraction, $\mathcal{B}$, for a lossy partial BSM versus the average number of signal-idler pairs per SPDC island per pump pulse, $G-1$, for (top to bottom) $\eta_T = 1,0.9, 0.8, \ldots, 0.5$.  It shows the high premium placed on the having $\eta_T \sim 1$ to maintain a high Bell-state fraction at the partial-BSM's output.  Indeed, whereas for lossless operation $G-1 = 0.5$ resulted in quasi-deterministic ($\Pr(\mathcal{H}_{\rm true}) > 0.25$) performance with unit Bell-state fidelity and unit Bell-state fraction, when $\eta_T = 0.9$ we need $G-1 = 0.0129$ to realize $\mathcal{B} = 0.99$.  Although that Bell-state fraction is accompanied by a unit Bell-state fidelity, it takes a wildly unrealistic 1381 islands in order to reach $\Pr(\mathcal{H}_{\rm true}) > 0.25$ with same-island heralding when $G-1 = 0.0129$.  

The explosive growth of the number of islands needed to achieve $\mathcal{B} \ge 0.99$ and $\mathcal{F} = 1$ with same-island heralding that is incurred in going from $\eta_T = 1$ to $\eta_T = 0.9$ can be avoided by employing cross-island together with same-island heralding.  We call such a heralding scheme same-plus-cross-island (SPCI) heralding.  Same-island heralding requires detection of an $H$-polarized photon and a $V$-polarized photon from the same island.  Thus same-island heralding offers only $N_I$ island-choice possibilities.  Because, as shown in Eq.~(\ref{islands_state}), the states generated by each island are iid TMSV states, detecting an $H$-polarized photon from the $n$th island and a $V$-polarized photon from the $m$th island, with $m\neq n$, also provides a suitable herald.  Now, with SPCI heralding, Alice and Bob need to be told which island had the $H$ detection and which had the $V$ detection, and their QRXs must implement  polarization-dependent frequency conversion in their mode-conversion modules, i.e., the $H$-polarized and $V$-polarized components of Alice and Bob's received photons have their respective center frequencies converted from the $n$th and $m$th islands' center frequencies to the center frequency needed for their quantum memories.  

With both same-island and cross-island heralds being employed the probability of an $n$th-island, $m$th-island herald---regardless of whether $m\neq n$ or $m=n$---is~\cite{footnoteD}
\begin{equation}
\Pr(\mathcal{H}_{nm}) = \frac{4[\eta_T(G-1)]^2}{[\eta_T(G-1)+1]^6},\mbox{ for $n,m = 1,2,\ldots, N_I$},
\label{nm_herald}
\end{equation}
with half of them being false heralds.  Continuing our assumption of allowing at most one herald per pump pulse, the per-pump-pulse probability of a true herald when both same-island and cross-island 
heralds are used is then~\cite{footnote6}
\begin{align}
\Pr(\mathcal{H}_{\rm true}) &= \{1 - 2[1-\sqrt{\Pr(\mathcal{H}_{nm})}\,]^{N_I} \nonumber \\[0.05in]
& + [1-\sqrt{\Pr(\mathcal{H}_{nm})}\,]^{2N_I}\}/2,
\label{true-herald2}
\end{align}
because there are now $N_I^2$ island-pair choices for the $H$ and $V$ detections but their heralding behaviors are \emph{not} statistically independent.  Consequently $\Pr(\mathcal{H}_{\rm true})$ with SPCI heralding scales approximately like $N_I^2$ only for $G-1 \ll 1$. Nevertheless, using same-island and cross-island heralding, 55 islands suffice to get $\Pr(\mathcal{H}_{\rm true}) > 0.25$ with $\mathcal{B} = 0.99$ and $\mathcal{F} = 1$ from a lossy partial BSM when $\eta_T = 0.9$, and 120 islands suffice to achieve that same performance when $\eta_T = 0.8$.  

\subsection{Islands-based ZALM with partial-BSM loss and propagation loss \label{lossy_prop}}
The final task in this paper is to study the effects of signal-path losses on islands-based ZALM.  In the previous subsection we presumed the ZALM QTX's signal paths were lossless.  Taking signal-path losses to be symmetrically distributed, they can be accounted for by subjecting the QTX's $n$th-island heralded outputs to sub-unit transmissivities $\eta_R$.  Furthermore, with the assumed source-in-the-middle architecture, it is not unreasonable---at least for an initial study---to lump together the ZALM QTX's signal path losses with symmetrically-distributed propagation losses en route to Alice and Bob's QRXs.  That case is considered in Appendix~\ref{AppendB}, where beam-splitter mode transformations, with transmissivity $\eta_R$, are first used to obtain the anti-normally-ordered characteristic function for the $S_A$ and $S_B$ modes, i.e., the modes arriving at Alice and Bob's QRXs.  Appendix~\ref{AppendB} then parallels what was done in Appendix~\ref{AppendA} to arrive at the following results for ZALM's Bell-state probabilities, given an $I'_{+H}I'_{-V}$ herald, with partial-BSM loss and propagation loss:  
\begin{widetext}
\begin{equation}
{}_{\tilde{\bf S}}\langle \psi^-|\hat{\rho}_{\tilde{\bf S}\mid I'_{+H}I'_{-V}}|\psi^-\rangle_{\tilde{\bf S}}  = \frac{N_S'^4}{2}\!\left[2(1-N_S')^2 -\frac{2\eta_R(3N_S'^3-5N_S'^2+2N_S')}{N_S} + \frac{\eta_R^2(5N_S'^4-6N_S'^3+2N_S'^2)}{N_S^2}\right],
\label{PrSnLossProp}
\end{equation}
and 
\begin{align}
{}_{\tilde{\bf S}}\langle \psi^+|\hat{\rho}_{\tilde{\bf S}\mid I'_{+H}I'_{-V}}|\psi^+\rangle_{\tilde{\bf S}}  &=
{}_{\tilde{\bf S}}\langle \phi^\pm|\hat{\rho}_{\tilde{\bf S}\mid I'_{+H}I'_{-V}}|\phi^\pm\rangle_{\tilde{\bf S}} 
\nonumber \\[.05in]
& = \frac{N_S'^4}{2}\!\left[2(1-N_S')^2 -\frac{2\eta_R(3N_S'^3-5N_S'^2+2N_S')}{N_S} + \frac{\eta_R^2(4N_S'^4-6N_S'^3+2N_S'^2)}{N_S^2}\right],
\label{PrEnLossProp}
\end{align}
\end{widetext}
where $\tilde{\bf S}$ denotes the $S_AS_B$ signal modes arriving at Alice and Bob's QRXs, and $N_S' \equiv [\eta_R/N_S + (1-\eta_R)]^{-1}.$

\begin{figure}[hbt]
    \centering
\includegraphics[width=0.99\columnwidth]{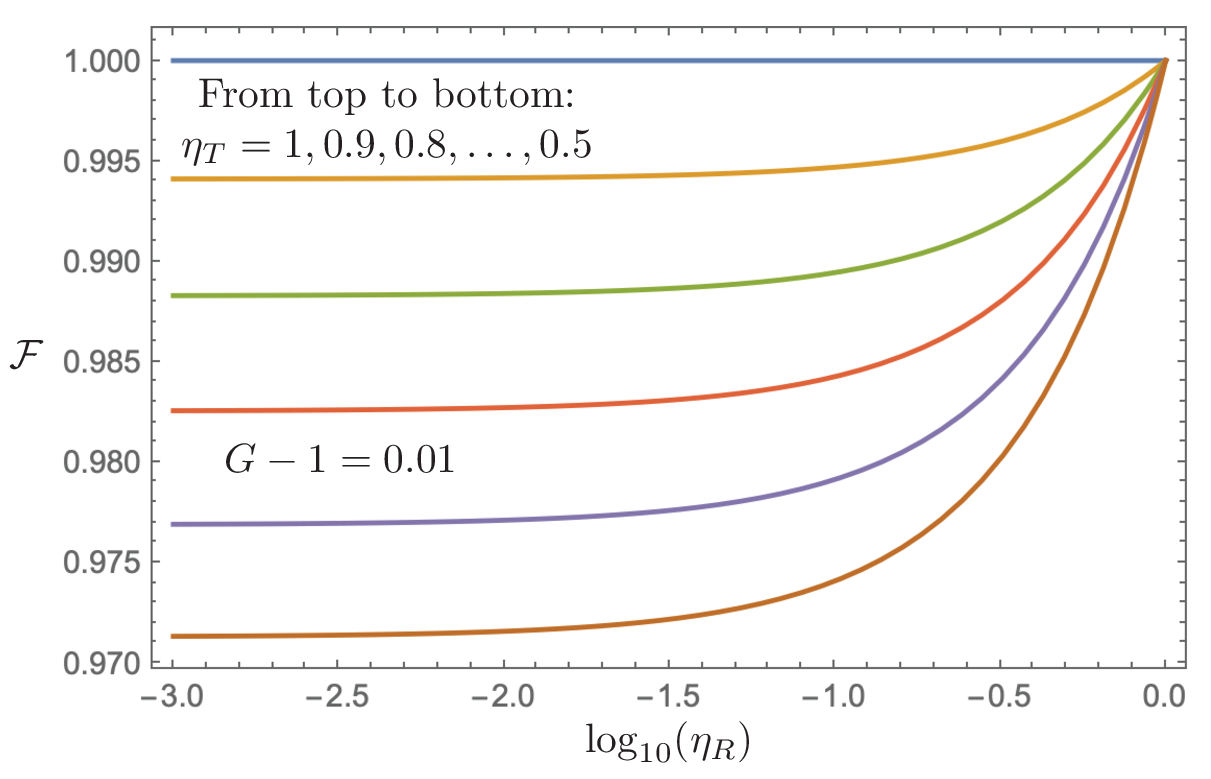}
    \caption{Performance of islands-based ZALM with both partial-BSM loss and propagation loss: Bell-state fidelity, $\mathcal{F}$, of the $S_AS_B$ state versus $\log_{10}(\eta_R)$ for $G-1 = 0.01$ and (top to bottom) $\eta_T= 1,0.9,0.8,\ldots,0.5.$ \label{BellFidelityR1} }
\end{figure}

\begin{figure}[hbt]
    \centering
\includegraphics[width=0.99\columnwidth]{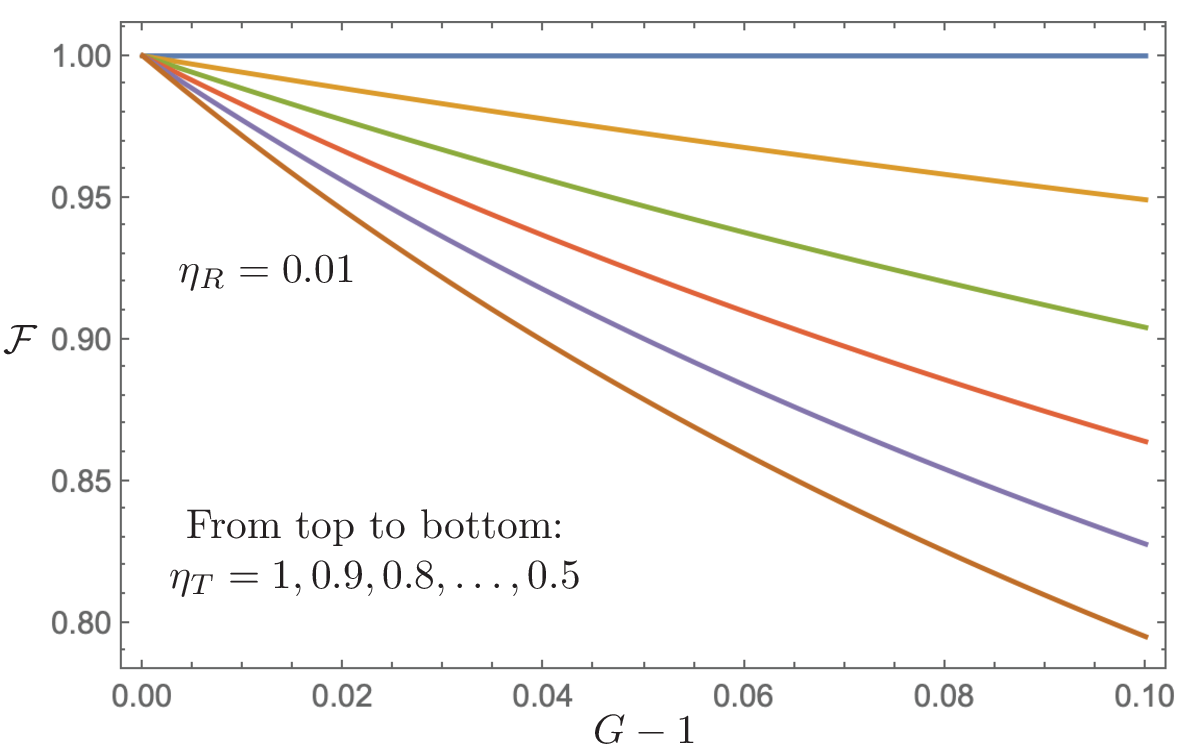}
    \caption{Performance of islands-based ZALM with both partial-BSM loss and propagation loss: Bell-state fidelity, $\mathcal{F}$, of the $S_AS_B$ state versus average number of signal-idler pairs per SPDC island per pump pulse, $G-1$, for $\eta_R = 0.01$ and (top to bottom) $\eta_T= 1,0.9,0.8,\ldots, 0.5.$ \label{BellFidelityR2} }
\end{figure}

From Eqs.~(\ref{PrSnLossProp}) and (\ref{PrEnLossProp})---and the fact that the corresponding results for the other three herald possibilities coincide with these $I'_{+H}I'_{-V}$-herald results---we get the Bell-state fidelity curves in Figs.~\ref{BellFidelityR1} and \ref{BellFidelityR2} when there is both partial-BSM loss and propagation loss.   Figure~\ref{BellFidelityR1} shows that $G-1 = 0.01$ and $\eta_T \ge 0.5$ results in Bell-state fidelities exceeding 0.97 for QTX-to-QRX losses as high as 30\,dB.  Figure~\ref{BellFidelityR2} shows that $G-1 = 0.0173$ gives $\mathcal{F} = 0.99$ when $\eta_T = 0.9$ and $\eta_R = 0.01$, while $G-1 = 8.59\times 10^{-3}$ does the same when $\eta_T = 0.8$ and $\eta_R = 0.01$.  

\begin{widetext}
Next on our agenda is evaluating islands-based ZALM's Bell-state fraction when there is both partial-BSM loss and propagation loss. From Eqs.~(\ref{PrSnLossProp}) and (\ref{PrEnLossProp}) we have
\begin{equation}
\Pr({\rm Bell}\mid I'_{+H}I'_{-V}) = 2N_S'^4\left[2(1-N_S')^2 -\frac{2\eta_R(3N_S'^3-5N_S'^2+2N_S')}{N_S} + \frac{\eta_R^2(4N_S'^4-6N_S'^3+2N_S'^2)}{N_S^2}\right]   + \frac{\eta_R^2N^{\prime 8}_S}{2N_S^2},
\end{equation} 
and from Appendix~\ref{AppendB} we have 
\begin{equation}
\Pr({\rm loadable}\mid I'_{+H}I'_{-V}) = 1-2N_S'^2\left[1-\frac{\eta_RN_S'}{2N_S}\right]^2  + N_S'^4\!\left[1-\frac{\eta_RN_S'}{N_S}\right]^2.
\label{loadable2}
\end{equation}
\end{widetext}

\begin{figure}[hbt]
    \centering
\includegraphics[width=0.99\columnwidth]{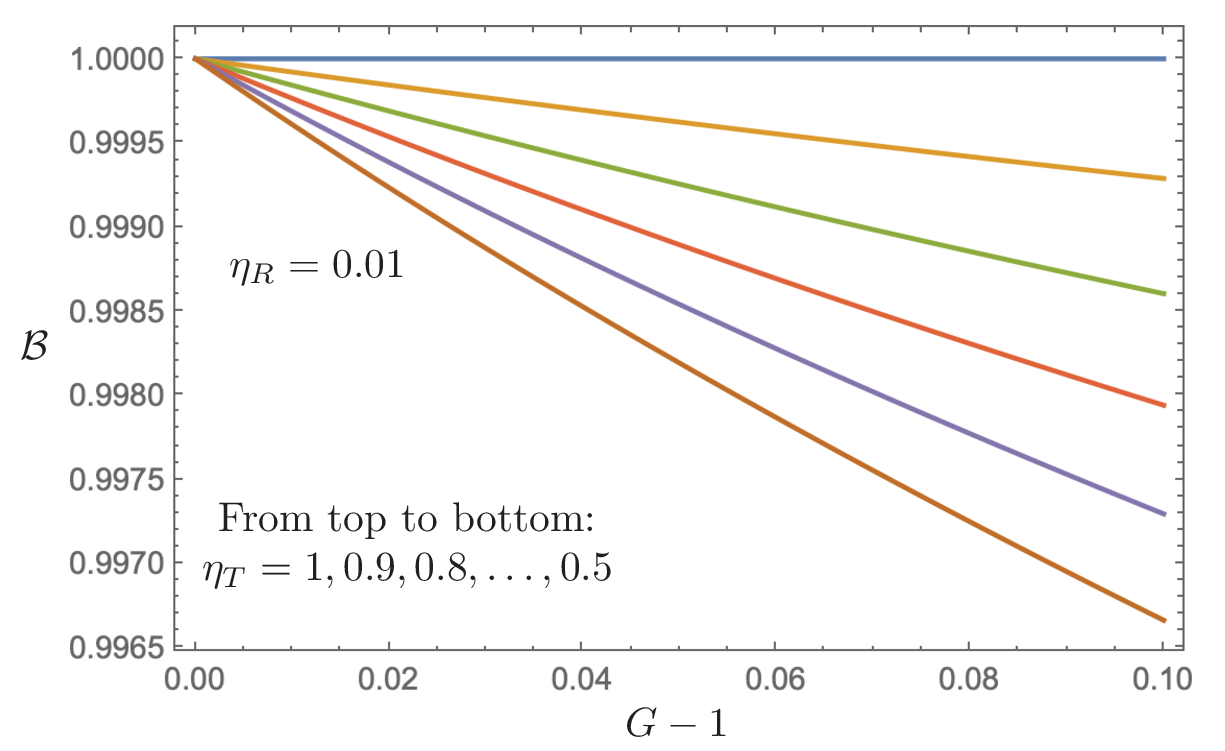}
    \caption{Performance of islands-based ZALM with both partial-BSM loss and propagation loss: Bell-state fraction, $\mathcal{B}$, of the $S_AS_B$ state versus average number of signal-idler pairs per SPDC island per pump pulse, $G-1$, for $\eta_R = 0.01$ and (top to bottom) $\eta_T= 1,0.9,0.8,\ldots, 0.5.$ \label{BellFractionR1} }
\end{figure}

\begin{figure}[hbt]
    \centering
\includegraphics[width=0.99\columnwidth]{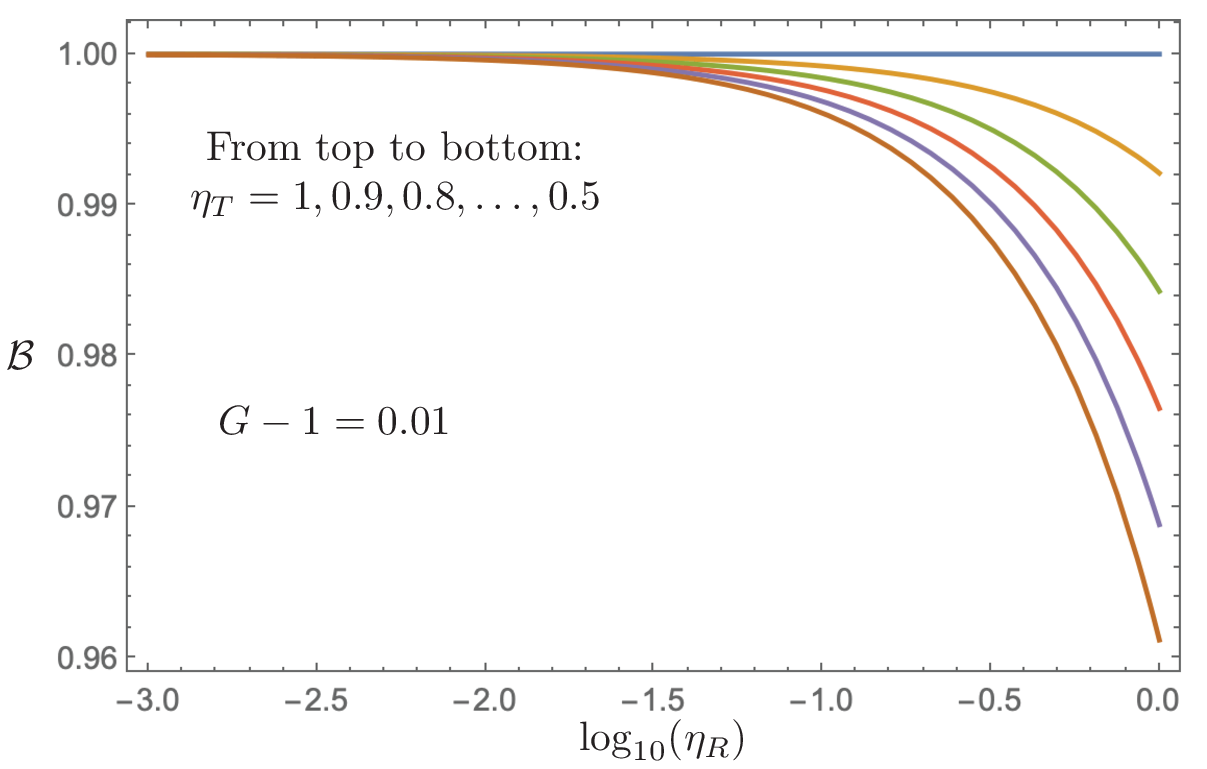}
    \caption{Performance of islands-based ZALM with both partial-BSM loss and propagation loss: Bell-state fraction, $\mathcal{B}$, of the $S_AS_B$ state versus $\log_{10}(\eta_R)$ for $G-1 = 0.01$ and (top to bottom) $\eta_T= 1,0.9,0.8.,\ldots,0.5.$ \label{BellFractionR2} }
\end{figure}

The preceding results also apply to the other three herald possibilities and lead to the Bell-state fraction behaviors in Figs.~\ref{BellFractionR1} and \ref{BellFractionR2}.  They show that very lossy propagation with $G-1 \le 0.1$ gives exceedingly high Bell-state fraction for $\eta_T \ge 0.5$, and for $G-1 = 0.01$ the Bell-state fraction increases as $\eta_R$ decreases.  Let us see what this behavior, combined with what we have already exhibited for the Bell-state fidelity, has to say about delivering high-quality entanglement through very lossy propagation with quasi-deterministic per-pump-pulse probability of a true herald.  

We noted earlier that $G-1 = 0.0173$ gives $\mathcal{F} = 0.99$ when $\eta_T = 0.9$ and $\eta_R = 0.01$, and $G-1 = 8.59 \times 10^{-3}$ does the same when $\eta_T = 0.8$ and $\eta_R = 0.01$.  We can now say that $G-1 = 0.0173$ gives $1-\mathcal{B} = 1.36\times 10^{-4} $ when $\eta_T = 0.9$ and $\eta_R = 0.01$, and $G-1 = 8.59 \times 10^{-3}$ does the same when $\eta_T = 0.8$ and $\eta_R = 0.01$.  In simple terms, these average numbers of signal-idler pairs per island per pump pulse result in delivery of an $S_AS_B$ state---from a lossy partial BSM through very lossy propagation---that is essentially confined to the Bell states' Hilbert space.  Furthermore, with SPCI heralding $\Pr(\mathcal{H}_{\rm true})$ exceeds 0.25 when $N_I = 41$ for $\eta_T = 0.9$, and $N_I = 91$ for $\eta_T = 0.8$ when $\eta_R = 0.01$ and we use the $G-1$ values given above.  So, whereas Chen~\emph{et al}.~\cite{Chen2023} used 800 channels to achieve high-fidelity, quasi-deterministic operation with same-channel heralding, we have demonstrated that \emph{far} fewer phase-matched spectral islands are needed to do the same with SPCI heralding.  

Some additional results from Appendix~\ref{AppendB} deserve mention at this point.  There we show that 
\begin{equation}
{}_{\tilde{\bf S}}\langle \psi^+|\hat{\rho}_{\tilde{\bf S}\mid I'_{+H}I'_{-V}}|\psi^-\rangle_{\tilde{\bf S}} = 0,
\end{equation}
and
\begin{equation}
{}_{\tilde{\bf S}}\langle \phi^\pm|\hat{\rho}_{\tilde{\bf S}\mid I'_{+H}I'_{-V}}|\psi^-\rangle_{\tilde{\bf S}} = 0,
\end{equation}
implying that, given an $I'_{+H}I'_{-V}$ herald, the projection of $\hat{\rho}_{\tilde{\bf S}\mid I'_{+H}I'_{-V}}$ into $\mathcal{H}_B \times \mathcal{H}_B$, where $\mathcal{H}_B$ is the $S_AS_B$ Bell-state Hilbert space, is diagonal in the Bell basis.  As explained in Appendix~\ref{AppendB}, the same Bell-state diagonality occurs for the other possible heralds.  The near-unit Bell-state fractions of the previous paragraph's examples then let us say that 
\begin{align}
\hat{\rho}_{\tilde{\bf S}\mid \psi^\pm} &\approx K^{-1}\!\left[\Pr(|\psi^\pm\rangle_{\tilde{\bf S}} \mid \psi^\pm)|\psi^\pm\rangle_{\tilde{\bf S}}\,{}_{\tilde{\bf S}}\langle\psi^{\pm}|\right. \nonumber \\[.05in]
&+ \Pr(|\psi^\mp\rangle_{\tilde{\bf S}} \mid \psi^\pm)|\psi^\mp\rangle_{\tilde{\bf S}}\,{}_{\tilde{\bf S}}\langle\psi^{\mp}| \nonumber \\[.05in]
&+ \left.\sum_{s=+,-} \Pr(|\phi^s\rangle_{\tilde{\bf S}} \mid \psi^\pm)|\phi^s\rangle_{\tilde{\bf S}}\,{}_{\tilde{\bf S}}\langle\phi^s|\right].
\label{BellStateProjection}
\end{align}
Here: $\Pr(\cdot\mid \psi^\pm)$ denotes a conditional probability given a $\psi^\pm$ herald; $\Pr(|\psi^\pm\rangle_{\tilde{\bf S}} \mid \psi^\pm)$ is given by the right-hand side of Eq.~(\ref{PrSnLossProp}); $\Pr(|\psi^\mp\rangle_{\tilde{\bf S}} \mid \psi^\pm)$ and $\Pr(|\phi^+\rangle_{\tilde{\bf S}} \mid \psi^\pm) = \Pr(|\phi^-\rangle_{\tilde{\bf S}} \mid \psi^\pm)$ are given by the right-hand side of Eq~(\ref{PrEnLossProp}); and  
\begin{equation}
K \equiv \sum_{s = +,-}\left[\Pr(|\psi^s\rangle_{\tilde{\bf S}} \mid \psi^\pm) + \Pr(|\phi^s\rangle_{\tilde{\bf S}} \mid \psi^\pm)\right]
\end{equation}
is a normalization constant.  

Note that the right-hand side of Eq.~(\ref{BellStateProjection}) is an \emph{exact} result for $\hat{\rho}_{\tilde{\bf S}_B\mid \psi^\pm}$, the projection of $\hat{\rho}_{\tilde{\bf S}\mid \psi^\pm}$ on to $\mathcal{H}_B\times \mathcal{H}_B$, regardless of the Bell-state fraction.  Thus, we can find the $S_AS_B$ states' Bell-state purity, $\mathcal{P} = {\rm Tr}(\hat{\rho}^2_{\tilde{\bf S}_B\mid \psi^\pm})$, from Eq.~(\ref{BellStateProjection}).  When $\mathcal{B} \approx 1$, this result becomes the full state's purity.  

\begin{figure}[hbt]
    \centering
\includegraphics[width=0.99\columnwidth]{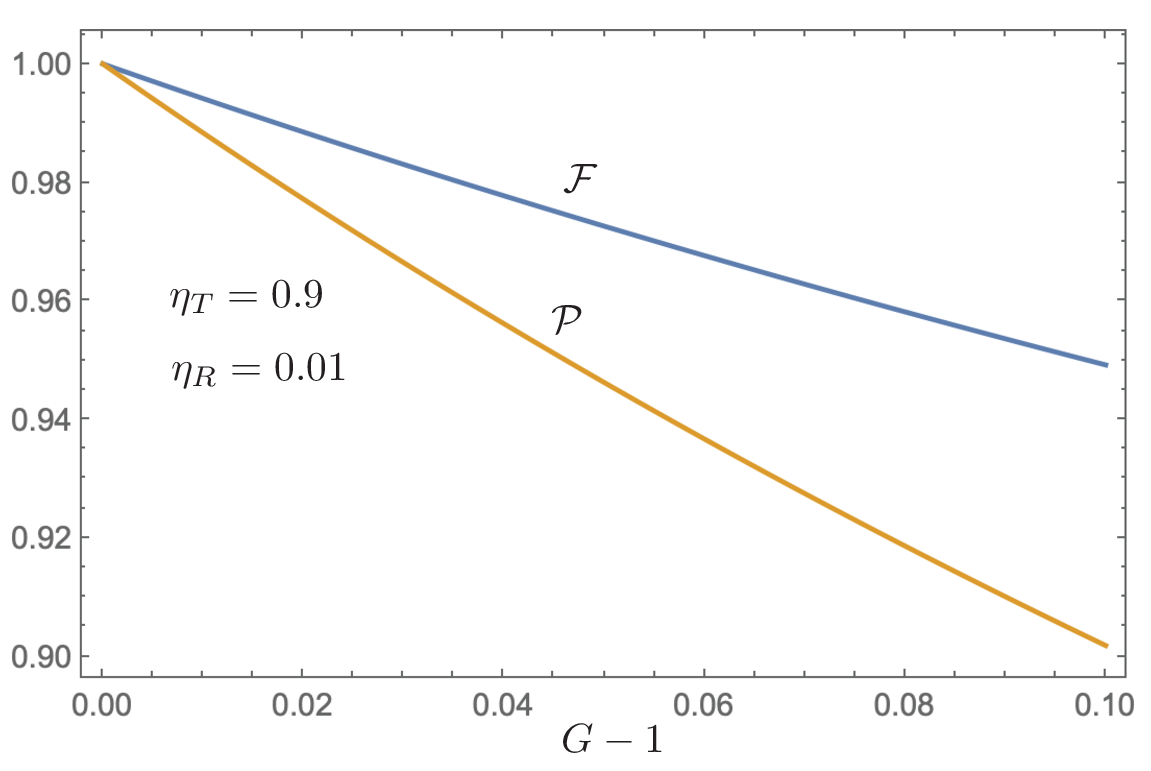}
    \caption{Performance of islands-based ZALM with both partial-BSM loss and propagation loss: Bell-state fidelity, $\mathcal{F}$, and Bell-state purity, $\mathcal{P}$, of the $S_AS_B$ state versus average number of signal-idler pairs per SPDC island per pump pulse, $G-1$, for $\eta_T = 0.9$ and $\eta_R= 0.01$ \label{FidelityPurity} }
\end{figure}

Figure~\ref{FidelityPurity} plots the Bell-state fidelity and the Bell-state purity versus the average number of signal-idler pairs per island per pump pulse for $\eta_T = 0.9$ and $\eta_R = 0.01.$  We see that fidelity is greater than the purity---$\mathcal{F} = 0.99$ versus $\mathcal{P} = 0.98$ when $G-1 = 0.0173$---something that we had previously seen for lossless operation of DWDM-channelized ZALM~\cite{Shapiro2024}.

\section{Conclusions and Discussion \label{discussion}}
Chen~\emph{et al}.~\cite{Chen2023} proposed ZALM as a means to avoid the ill-effects of switch losses at the source in multiplexing SPDCs, and thus achieve high-fidelity ($>$\,99\%), quasi-deterministic ($\ge$\,25\% per-pump-pulse probability) entanglement generation.  Their proposal relies on a burst of pump pulses to generate a striped frequency-domain biphoton wave function over a very broad (10\,THz) bandwidth that, together with narrowband (1\,GHz) filtering in the partial BSM and suitable (12.5\,GHz) channelization at the receivers, leads to the need for a large number (800) of channels whose frequency-domain biphoton wave functions each approximate that of a spectrally factorable, i.e., single-temporal-mode, state.  Chen~\emph{et al}.'s ZALM comes, however, with a heavy technological burden in that an 800-channel system requires a partial-BSM apparatus containing 3200 high quantum-efficiency SPDs, which would typically be cryo-cooled superconducting-nanowire single-photon detectors.  Temporal multiplexing of the detection process could reduce the number of required SPDs, but that comes at the expense of reducing the maximum pump-pulse rate that can be employed.  (In that regard, note that Chen~\emph{et al}.\@ had assumed a $3\times 10^{10}\,{\rm s}^{-1}$ pump-pulse rate.)  Furthermore, Chen~\emph{et al}.'s analysis did not fully account for multipair events and losses in the partial BSM and in propagation from the ZALM QTX to Alice and Bob's QRXs.    

In seeking a better alternative for the ZALM QTX, as well as providing a more comprehensive analysis, we drew inspiration from Morrison~\emph{et al}.'s demonstration of an SPDC producing a biphoton wave function comprised of 8 near-ideal single-temporal-mode spectral islands~\cite{Morrison2022}.  Thus, we assumed SPDCs whose nonlinear crystals had $N_I$ ideal spectral islands, and then derived the anti-normally-ordered characteristic functions for the output signals' state from a lossy partial BSM, and for the joint state delivered, after lossy propagation, to Alice and Bob's QRXs.   From these characteristic functions we obtained analytic results for the Bell-state fidelities and Bell-state fractions of the output signals' state and the joint state arriving at Alice and Bob's QRXs.  Then, assuming at most one herald is sent to Alice and Bob per pump pulse and SPCI heralding, we found that 99\% Bell-state fidelity, $>$\,99.98\% Bell-state fraction, and $\ge$\,25\% per-pump-pulse probability of a true herald can be obtained with 41 islands if the partial BSM is 90\% efficient, and with 91 islands if the partial BSM is 80\% efficient.  If the number of islands is limited by practical considerations to, for example, 20, then a 95\% efficient partial BSM will give 99\% Bell-state fidelity, $>$\,99.98\% Bell-state fraction, and $\ge$\,25\% per-pump-pulse probability of a true herald.

The rate, $\mathcal{R}_E$, at which islands-based ZALM delivers heralded polarization-entangled photon pairs to Alice and Bob is this paper's final takeaway.  It is given by~\cite{footnote7}
\begin{equation}
\mathcal{R}_E = \mathcal{R}_P \Pr(\mathcal{H}) \Pr(|\psi^\pm\rangle_{\tilde{\bf S}} \mid \psi^\pm),
\end{equation}
where $\mathcal{R}_P$ is the pump's pulse rate, and $\Pr(\mathcal{H}) = 2\Pr(\mathcal{H}_{\rm true})$, the per-pump-pulse herald probability, includes both true and false heralds.  Assuming SPCI heralding with at most one herald sent to Alice and Bob per pump pulse, and $R_P = 10^{10}\,{\rm s}^{-1}$, we get $R_E = 2.54\times 10^5\,{\rm s}^{-1}$ when $\eta_T = 0.9$, $G-1 = 0.0173$, $\eta_R = 0.01$, and $N_I = 41$.  Reducing $\eta_T$ to 0.8 and $G-1$ to $8.59\times 10^{-3}$, while increasing $N_I$ to 91, results in an equivalent delivery rate of $\mathcal{R}_E = 2.53 \times 10^5\,{\rm s}^{-1}$.  Alternatively, improving the partial-BSM's efficiency to $\eta_T = 0.95$, and increasing pump strength to get $G-1 = 0.0353$, gives $\mathcal{R}_E = 2.57 \times 10^5\,{\rm s}^{-1}$ from $N_I = 20$ islands. All of these examples provide 99\% Bell-state fidelity and $>$\,99.98\% Bell-state fraction in the presence of propagation loss commensurate with: (a) 100\,km fiber connections between the ZALM QTX and Alice and Bob's QRXs; or (b) 1500-km-long satellite-to-ground free-space (no absorption, scattering, or turbulence) links at 1.55\,$\upmu$m wavelength between a ZALM QTX with 10-cm-diameter optics and Alice and Bob's QRXs with 1-m-diameter optics.  

These preceding rates are admittedly optimistic, but, because they are orders-of-magnitude higher than current state of the art \emph{and} with lower equipment burden than DWDM-channelized ZALM, they strongly suggest continued study of islands-based ZALM.   There are a variety of theoretical directions in which such follow-on work could be pursued.  One is a more comprehensive exploration of the Bell-state fidelity and Bell-state fraction as functions of $G-1$, $\eta_T$, $\eta_R$, and the available $N_I$.  Such an exploration, in conjunction with a capabilities assessment of domain engineering PPLN, could set realistic goals for an initial islands-based ZALM source.

Another theory area to study is the Duan-Kimble loading~\cite{Duan2004} of Alice and Bob's pair of intracavity color-center quantum memories.  Islands-based ZALM affords an extremely high Bell-state fraction, permitting us to assume that Bell states are the only loadable states arriving at Alice and Bob's QRXs.  For Bell-state illumination we can use results from Raymer~\emph{et al}.~\cite{Raymer2024} and Shapiro~\emph{et al}.~\cite{Shapiro2024} to obtain the density operator for the loaded memories. In particular, Shapiro~\emph{et al}.\@ already showed that narrowband operation---in which the ZALM QRXs' bandwidth-compression module is such that the quantum memories' state-dependent reflectivities are constant across the photon bandwidth---leads to a perfect transfer of the incoming biphoton state from DWDM-channelized ZALM to the memories.  It turns out that such is not the case for narrowband loading of the Bell state received from islands-based ZALM unless the memories are completely lossless~\cite{footnote8}. Nevertheless, Shapiro~\emph{et al.}'s loading analysis establishes a route to quantifying islands-based ZALM's Duan-Kimble loading both inside and outside the narrowband regime.   

The just-mentioned follow-on studies are natural progressions from Chen~\emph{et al}., Raymer~\emph{et al.}, Shapiro~\emph{et al}., and the present paper.  A somewhat different tack, but nonetheless an important one, is to consider replacing the Sagnac sources and partial BSM with a new alternative,  proposed by Chahine~\emph{et al}.~\cite{Chahine2025}, for heralded production of polarization-entangled Bell states.  Chahine~\emph{et al}.'s proposal uses only two SPDCs and has a simpler heralding procedure than the dual-Sagnac arrangement considered here.  For islands-based ZALM it then offers the possibility of lower loss in the heralding, which could increase the $G-1$ value at which  high Bell-state fidelity and high Bell-state fraction are obtained from which the number of islands needed for quasi-deterministic operation could be reduced.  That said, the Chahine proposal's joint density operator for the signal states sent to Alice and Bob does not match what we have derived in this paper unless the ZALM QTX is lossless, so an analysis of its performance metrics is needed to see whether it truly offers an improvement over what we have found in this paper for the dual-Sagnac source.  

\begin{center}
\begin{table}
\begin{tabular}{|c||c|}\hline\hline 
Symbol & Meaning  \\ \hline
$G-1$ & average photon number per SPDC island \\  
$N _I$ & number of islands per SPDC \\ 
$\mathcal{H}_n$ & $n$th same-island heralding event \\
$\mathcal{H}_{nm}$ & $nm$th cross-island heralding event \\ 
$\mathcal{H}_{\rm true}$ & true heralding event for ZALM source \\ 
$\eta_T$ & transmitter-loss transmissivity \\
$\hat{\rho}$ & normalized density operator \\
$\tilde{\rho}_{\cdot \mid \cdot}$ & unnormalized conditional density operator\\ 
$\hat{\rho}_{\cdot \mid \cdot}$ & normalized conditional density operator\\ 
$\chi_A^\rho(\cdot)$ & anti-normally ordered characteristic function\\ 
$N_S$ & $[\eta_T(G-1)+1]/G$ \\ 
$\mathcal{B}$ & Bell-state fraction \\
$\mathcal{F}$ & Bell-state fidelity \\ 
$\eta_R$ & propagation-loss transmissivity \\
$N_S'$ & $[\eta_R/N_S + (1-\eta_R)]^{-1}$ \\
$\mathcal{H}_B$ & Bell-state Hilbert space \\
$\mathcal{P}$ & Bell-state purity \\
$\mathcal{R}_P$ & SPDC pump-pulse rate \\
$\mathcal{R}_E$ & entanglement delivery rate\\ 
\hline \hline
\end{tabular}
\caption{Glossary of principal symbols.}
\end{table}
\end{center}

\acknowledgments
This work was supported by the Engineering Research Centers Program of the National Science Foundation
under Grant \#1941583 to the NSF-ERC Center for Quantum Networks.  The authors thank Franco Wong for encouragement and advice in our pursuit of a ZALM source capable of high-purity, quasi-deterministic entanglement distribution, and we thank Yousef Chahine for alerting us to his new concept for a heralded source of polarization-entangled Bell states.
 
 \appendix
 
 \section{Islands-based ZALM partial-BSM loss:  Derivations \label{AppendA}}
In this appendix we derive the performance metrics for islands-based ZALM with a lossy partial-BSM apparatus, i.e., one whose inefficiencies have been lumped together into a sub-unit quantum efficiency, $\eta_T$, for each of its SPDs.
The initial quantities of interest are the conditional Bell-state output probabilities from Eqs.~(\ref{s_nIV.a}) and (\ref{e_nIV.a}) given an $I'_{+H}I'_{-V}$ herald has occurred.  We begin by reworking Eq.~(\ref{chiAstart})'s expression for the anti-normally-ordered characteristic function associated with the joint signal-idler density operator, $\hat{\rho}_{\bf SI}$, of a single island.

As a first step, it is easily seen that Eq.~(\ref{chiAstart}) factors into
\begin{align}
\chi&_A^{\rho_{\bf SI}}(\bzeta) = \exp\!\left[-G\bzeta^{(R)T}\bzeta^{(R)} + 2\sqrt{G(G-1)}\,\bzeta^{(R)T}_S\bzeta^{(R)}_I\right] \nonumber \\[.05in]
&\times \exp\!\left[-G\bzeta^{(I)T}\bzeta^{(I)} - 2\sqrt{G(G-1)}\,\bzeta^{(I)T}_S\bzeta^{(I)}_I\right],
\label{chiAstep1}
\end{align}
where $\bzeta^{(R)}$ and $\bzeta^{(I)}$ are the real and imaginary parts of $\bzeta$.  To proceed further, we rewrite the exponents in Eq.~(\ref{chiAstep1}) as
\begin{align}
-G\bzeta^{(R)T}\bzeta^{(R)} +\, & 2\sqrt{G(G-1)}\,\bzeta^{(R)T}_S\bzeta^{(R)}_I \nonumber \\[.05in] &= -
\bzeta^{(R)T}\tilde{\Lambda}_{\bf SI}^{(R)}\bzeta^{(R)}/2,
\end{align}
and
\begin{align}
-G\bzeta^{(I)T}\bzeta^{(I)} -\, & 2\sqrt{G(G-1)}\,\bzeta^{(I)T}_S\bzeta^{(I)}_I \nonumber \\[.05in] &= -
\bzeta^{(I)T}\tilde{\Lambda}_{\bf SI}^{(I)}\bzeta^{(I)}/2,
\end{align}
where
\begin{equation}
\tilde{\Lambda}_{\bf SI}^{(R)} = \left[\begin{array}{cc} \tilde{\Lambda}_{SS} & \tilde{\Lambda}_{SI} \\[.05in] \tilde{\Lambda}_{SI}^T & \tilde{\Lambda}_{II}\end{array}\right],
\end{equation}
and 
\begin{equation}
\tilde{\Lambda}_{\bf SI}^{(I)} = \left[\begin{array}{cc} \tilde{\Lambda}_{SS} & -\tilde{\Lambda}_{SI} \\[.05in] -\tilde{\Lambda}_{SI}^T & \tilde{\Lambda}_{II}\end{array}\right],
\end{equation}
with
\begin{equation}
\tilde{\Lambda}_{SS} =\tilde{\Lambda}_{II} =  2G\,\mathbb{I}_4,
\end{equation}
and
\begin{equation}
\tilde{\Lambda}_{SI} = -2\sqrt{G(G-1)}\,\mathbb{I}_4,
\end{equation}
for $\mathbb{I}_4$ being the 4D identity matrix.

Next, we introduce the beam-splitter relations that convert the ${\bf I}\equiv \{I_{1_H},I_{1_V},I_{2_H},I_{2_V}\}$ modes into the ${\bf I}_\pm\equiv \{I_{+H},I_{+V},I_{-H},I_{-V}\}$ modes, i.e.,
\begin{equation}
I_{\pm P} = \frac{I_{1_P} \pm I_{2_P}}{\sqrt{2}}, \mbox{ for $P = H,V$},
\end{equation}
that allow us to obtain the following anti-normally-ordered characteristic function for $\hat{\rho}_{{\bf SI}_\pm}$, viz.,
\begin{align}
\chi_A^{\rho_{{\bf SI}_\pm}}(\bzeta) &= \exp\!\left[-
\bzeta^{(R)T}\tilde{\Lambda}_{{\bf SI}_\pm}^{(R)}\bzeta^{(R)}/2\right] \nonumber \\[.05in]
& \times \exp\!\left[-
\bzeta^{(I)T}\tilde{\Lambda}_{{\bf SI}_\pm}^{(I)}\bzeta^{(I)}/2\right],
\end{align} 
where
\begin{equation}
\tilde{\Lambda}_{{\bf SI}_\pm}^{(R)} = \left[\begin{array}{cc} \tilde{\Lambda}_{SS} & \tilde{\Lambda}_{{SI}_\pm} \\[.05in] \tilde{\Lambda}_{{SI}_\pm}^T & \tilde{\Lambda}_{II}\end{array}\right],
\end{equation}
and 
\begin{equation}
\tilde{\Lambda}_{{\bf SI}_\pm}^{(I)} = \left[\begin{array}{cc} \tilde{\Lambda}_{SS} & -\tilde{\Lambda}_{{SI}_\pm} \\[.05in] -\tilde{\Lambda}_{{SI}_\pm}^T & \tilde{\Lambda}_{II}\end{array}\right],
\end{equation}
with
\begin{widetext}
\begin{equation}
\tilde{\Lambda}_{SI_\pm} = 
\left[\begin{array}{cccc} -\sqrt{2G(G-1)} & 0 & -\sqrt{2G(G-1)} & 0 \\[.05in]
0 & -\sqrt{2G(G-1)} & 0 & -\sqrt{2G(G-1)} \\[.05in]
-\sqrt{2G(G-1)} & 0 & \sqrt{2G(G-1)} & 0 \\[.05in]
0 & -\sqrt{2G(G-1)} & 0 & \sqrt{2G(G-1)} \end{array}\right].
\end{equation}
\end{widetext}

At this point we account for the SPDs' sub-unit quantum efficiency by recognizing that $\eta_T < 1$ detection of the ${\bf I}_\pm$ modes is equivalent to unit quantum efficiency detection of the ${\bf I}' \equiv \{I'_{+H},I'_{+V},I'_{-H},I'_{-V}\}$ modes whose annihilation operators are
\begin{equation}
\hat{a}'_{\pm P} = \sqrt{\eta_T}\,\hat{a}_{\pm P} +\sqrt{1-\eta_T}\,\hat{b}_{\pm P}, \mbox{ for $P=H,V,$}
\end{equation}
where the $\{\hat{b}_{\pm P}\}$ modes are in their vacuum states.  It now follows that the anti-normally-ordered characteristic function of $\hat{\rho}_{{\bf SI}'}$ is 
\begin{align}
\chi_A^{\rho_{{\bf SI}'}}(\bzeta) &= \exp\!\left[-
\bzeta^{(R)T}\tilde{\Lambda}_{{\bf SI}'}^{(R)}\bzeta^{(R)}/2\right] \nonumber \\[.05in]
& \times \exp\!\left[-
\bzeta^{(I)T}\tilde{\Lambda}_{{\bf SI}'}^{(I)}\bzeta^{(I)}/2\right],
\end{align} 
where
\begin{equation}
\tilde{\Lambda}_{{\bf SI}'}^{(R)} = \left[\begin{array}{cc} \tilde{\Lambda}_{SS} & \tilde{\Lambda}_{SI'} \\[.05in] \tilde{\Lambda}_{SI'}^T & \tilde{\Lambda}_{I'I'}\end{array}\right],
\end{equation}
and 
\begin{equation}
\tilde{\Lambda}_{{\bf SI}'}^{(I)} = \left[\begin{array}{cc} \tilde{\Lambda}_{SS} & -\tilde{\Lambda}_{SI'} \\[.05in] -\tilde{\Lambda}_{SI'}^T & \tilde{\Lambda}_{I'I'}\end{array}\right],
\end{equation}
with
\begin{widetext}
\begin{equation}
\tilde{\Lambda}_{SI'} = 
\left[\begin{array}{cccc} -\sqrt{2\eta_TG(G-1)} & 0 & -\sqrt{2\eta_TG(G-1)} & 0 \\[.05in]
0 & -\sqrt{2\eta_TG(G-1)} & 0 & -\sqrt{2\eta_TG(G-1)} \\[.05in]
-\sqrt{2\eta_TG(G-1)} & 0 & \sqrt{2\eta_TG(G-1)} & 0 \\[.05in]
0 & -\sqrt{2\eta_TG(G-1)} & 0 & \sqrt{2\eta_TG(G-1)} \end{array}\right],
\end{equation}
\end{widetext}
and
\begin{equation}
\tilde{\Lambda}_{I'I'} = 2[\eta_T(G-1) + 1]\mathbb{I}_4.
\end{equation}

We are almost but not quite ready to start evaluating the conditional Bell-state output probabilities for the case of an $I'_{+H}I'_{-V}$ herald.  What we need is to recast $\chi_A^{\rho_{{\bf SI}'}}(\bzeta)$ in a form that will permit 16D polynomial-weighted inverse Fourier transforms of it to be performed analytically.  Toward that end we define
\begin{equation}
\Lambda^{(R)}_{{\bf SI}'}\equiv \left(\tilde{\Lambda}^{(R)}_{{\bf SI}'}\right)^{-1} = 
\left[\begin{array}{cc} \Lambda_{SS} & \Lambda_{SI'} \\[.05in]
\Lambda_{SI'}^T & \Lambda_{I'I'}\end{array} \right],
\end{equation}
and
\begin{equation}
\Lambda^{(I)}_{{\bf SI}'}\equiv \left(\tilde{\Lambda}^{(I)}_{{\bf SI}'}\right)^{-1} = 
\left[\begin{array}{cc} \Lambda_{SS} & -\Lambda_{SI'} \\[.05in]
-\Lambda_{SI'}^T & \Lambda_{I'I'}\end{array} \right],
\end{equation}
where
\begin{equation}
\Lambda_{SS} = \frac{[\eta_T(G-1)+1]}{2G}\mathbb{I}_4,
\label{LamSS}
\end{equation}
\begin{equation}
\Lambda_{I'I'} = \mathbb{I}_4/2
\end{equation}
and
\begin{widetext}
\begin{equation}
\Lambda_{SI'} = 
\left[\begin{array}{cccc} \frac{\displaystyle \sqrt{\eta_TG(G-1)}}{\displaystyle 2\sqrt{2}\,G} & 0 & \frac{\displaystyle \sqrt{\eta_TG(G-1)}}{\displaystyle 2\sqrt{2}\,G} & 0 \\[.05in]
0 & \frac{\displaystyle \sqrt{\eta_TG(G-1)}}{\displaystyle 2\sqrt{2}\,G} & 0 & \frac{\displaystyle \sqrt{\eta_TG(G-1)}}{\displaystyle 2\sqrt{2}\,G}\\[.05in]
\frac{\displaystyle\sqrt{\eta_TG(G-1)}}{\displaystyle 2\sqrt{2}\,G} & 0 & -\frac{\displaystyle \sqrt{\eta_TG(G-1)}}{\displaystyle 2\sqrt{2}\,G} & 0 \\[.05in]
0 & \frac{\displaystyle \sqrt{\eta_TG(G-1)}}{\displaystyle 2\sqrt{2}\,G} & 0 & -\frac{\displaystyle \sqrt{\eta_TG(G-1)}}{\displaystyle 2\sqrt{2}\,G} \end{array}\right].
\end{equation}

The upshot of what we have just done is that $\chi_A^{\rho_{{\bf SI}'}}(\bzeta)$ is now given by
\begin{align}
\chi_A^{\rho_{{\bf SI}'}}(\bzeta) &= \exp\!\left[-
\bzeta^{(R)T}(\Lambda_{{\bf SI}'}^{(R)})^{-1}\bzeta^{(R)}/2\right] \exp\!\left[-
\bzeta^{(I)T}(\Lambda_{{\bf SI}'}^{(I)})^{-1}\bzeta^{(I)}/2\right] \\[.05in]
&= \left[(2\pi)^4\sqrt{{\rm det}(\Lambda_{{\bf SI}'}^{(R)})}\,p_{\bxi^{(R)}}(\bzeta^{(R)})\right]\left[(2\pi)^4\sqrt{{\rm det}(\Lambda_{{\bf SI}'}^{(I)})}\,p_{\bxi^{(I)}}(\bzeta^{(I)})\right],
\end{align}
where $p_{\bxi^{(R)}}(\bzeta^{(R)})$ and $p_{\bxi^{(I)}}(\bzeta^{(I)})$ are the probability density functions (pdfs) for 8D, real-valued, zero-mean Gaussian random vectors with covariance matrices $\Lambda_{{\bf SI}'}^{(R)}$ and $\Lambda_{{\bf SI}'}^{(I)}$, respectively.  Now, using standard results for Gaussian random vectors and ${\rm det}(\Lambda_{{\bf SI}'}^{(R)})={\rm det}(\Lambda_{{\bf SI}'}^{(I)}) = 1/256\,G^4$, we get our final rewrite of $\chi_A^{\rho_{{\bf SI}'}}(\bzeta)$:
\begin{align}
\chi_A^{\rho_{{\bf SI}'}}(\bzeta)&= (\pi^8/G^4)\,p_{\bxi_S^{(R)}}(\bzeta_S^{(R)})\,p_{\bxi_{I'}^{(R)}\mid\bxi_S^{(R)}}(\bzeta_{I'}^{(R)}\mid \bzeta_S^{(R)})\,p_{\bxi_S^{(I)}}(\bzeta_S^{(I)})\,p_{\bxi_{I'}^{(I)}\mid\bxi_S^{(I)}}(\bzeta_{I'}^{(I)}\mid \bzeta_S^{(I)}).
\label{chiAstep2} 
\end{align}
\end{widetext}
Here, $p_{\bxi_S^{(R)}}(\bzeta_S^{(R)})$ and $p_{\bxi_S^{(I)}}(\bzeta_S^{(I)})$ are the marginal pdfs of $\bxi_S^{(R)}$ and $\bxi_S^{(I)}$, i.e., 4D zero-mean, covariance-matrix $\Lambda_{SS}$, real-valued Gaussian random vectors.  On the other hand, $p_{\bxi_{I'}^{(R)}\mid\bxi_S^{(R)}}(\bzeta_{I'}^{(R)}\mid \bzeta_S^{(R)})$ and $p_{\bxi_{I'}^{(I)}\mid\bxi_S^{(I)}}(\bzeta_{I'}^{(I)}\mid \bzeta_S^{(I)})$ are, respectively, the conditional pdfs for $\bxi_{I'}^{(R)}$ given $\bxi_S^{(R)} = \bzeta_S^{(R)}$ and for  
$\bxi_{I'}^{(I)}$ given $\bxi_S^{(I)}= \bzeta_S^{(I)}$.  Those probability densities are 4D Gaussians with conditional mean vectors
\begin{equation}
\mathbb{E}(\bxi_{I'}^{(R)}\mid \bxi_S^{(R)} = \bzeta_S^{(R)}) = \Lambda_{SI'}^T\Lambda_{SS}^{-1}\bzeta_S^{(R)},
\end{equation}
and
\begin{equation}
\mathbb{E}(\bxi_{I'}^{(I)}\mid \bxi_S^{(I)} = \bzeta_S^{(I)}) = -\Lambda_{SI'}^T\Lambda_{SS}^{-1}\bzeta_S^{(I)},
\end{equation}
where
\begin{widetext}
\begin{equation}
\Lambda_{SI'}^T\Lambda_{SS}^{-1} =  \left[\begin{array}{cccc} \frac{\displaystyle \sqrt{\eta_TG(G-1)}}{\displaystyle \sqrt{2}\,[\eta_T(G-1)+1]} & 0 & \frac{\displaystyle \sqrt{\eta_TG(G-1)}}{\displaystyle \sqrt{2}\,[\eta_T(G-1)+1]} & 0 \\[.05in]
0 & \frac{\displaystyle \sqrt{\eta_TG(G-1)}}{\displaystyle \sqrt{2}\,[\eta_T(G-1)+1]} & 0 & \frac{\displaystyle \sqrt{\eta_TG(G-1)}}{\displaystyle \sqrt{2}\,[\eta_T(G-1)+1]} \\[.05in]
\frac{\displaystyle \sqrt{\eta_TG(G-1)}}{\displaystyle \sqrt{2}\,[\eta_T(G-1)+1]} & 0 & -\frac{\displaystyle \sqrt{\eta_TG(G-1)}}{\displaystyle \sqrt{2}\,[\eta_T(G-1)+1]} & 0 \\[.05in]
0 & \frac{\displaystyle \sqrt{\eta_TG(G-1)}}{\displaystyle \sqrt{2}\,[\eta_T(G-1)+1]} & 0 & -\frac{\displaystyle \sqrt{\eta_TG(G-1)}}{\displaystyle \sqrt{2}\,[\eta_T(G-1)+1]} 
\end{array}\right],
\end{equation}
\end{widetext}
and with conditional covariance matrices
\begin{align}
\Lambda^{(R)}_{I'I'\mid SS} &= \Lambda^{(I)}_{I'I'\mid SS} 
 =  \Lambda_{I'I'}-\Lambda_{SI'}^T\Lambda_{SS}^{-1}\Lambda_{SI'} \\[.05in] &= \frac{\mathbb{I}_4}{2[\eta_T(G-1)+1]}.
\end{align}
 
With the preceding $\chi_A^{\rho_{{\bf SI}'}}(\bzeta)$ results in hand we can now find $\hat{\rho}_{{\bf S}\mid I'_{+H}I'_{-V}}$, the density operator for the ZALM source's signal modes conditioned on there having been an $I'_{+H}I'_{-V}$ herald.  Once that is done, the desired conditional Bell-state output probabilities, viz., Eqs.~(\ref{s_nIV.a}) and (\ref{e_nIV.a}), are easily obtained.  Before proceeding, however, it is worth clarifying a point about the probability densities appearing in the equations above.  They do \emph{not} mean that the random vectors $\bxi^{(R)}$ and $\bxi^{(I)}$ have physical interpretations.  Instead, identifying terms in $\chi_A^{\rho_{{\bf SI}'}}(\bzeta)$ as Gaussian pdfs allows us to derive analytic performance results for islands-based ZALM with a lossy partial BSM.  These derivations are accomplished by means of Gaussian moment factoring, with each derivation in effect performing a 16D integral over a real-valued function.  

Now to work.  The unnormalized conditional density operator for the the ZALM source's signal modes, conditioned on there having been an $I'_{+H}I'_{-V}$ herald, is
\begin{align}
\tilde{\rho}&_{{\bf S}\mid I'_{+H}I'_{-V}} = \nonumber \\[.05in]
&{}_{I'_{-V}}\!\langle 1|\,{}_{I'_{-H}}\!\langle 0|\,{}_{I'_{+V}}\!\langle 0|\,{}_{I'_{+H}}\!\langle 1|\hat{\rho}_{{\bf SI}'}|1\rangle_{I'_{+H}}|0\rangle_{I'_{+V}}|0\rangle_{I'_{-H}}|1\rangle_{I'_{-V}}. 
\end{align}
Replacing $\hat{\rho}_{{\bf SI}'}$ with the operator-valued inverse Fourier transform of $\chi_A^{\rho_{{\bf SI}'}}(\bzeta)$, cf.\@ Eq.~(\ref{inverse_transform}), and evaluating the inner products we get
\begin{widetext}
\begin{equation}
\tilde{\rho}_{{\bf S}\mid I'_{+H}I'_{-V}} = \int\!\frac{{\rm d}^{16}\bzeta}{\pi^8}\,\chi_A^{\rho_{{\bf SI}'}}(\bzeta)(1-|\zeta_{I'_{+H}}|^2)(1-|\zeta_{I'_{-V}}|^2)
e^{-\bhata_S^\dagger\bzeta_S}e^{\bzeta_S^\dagger\bhata_S}.
\end{equation}
Using Eq.~(\ref{chiAstep2}), evaluating the expectations over the conditional probability densities, employing $\Lambda_{SS}$ from Eq.~(\ref{LamSS}), and coalescing terms we arrive at
\begin{equation}
\tilde{\rho}_{{\bf S}\mid I'_{+H}I'_{-V}} = \frac{[\eta_T(G-1)]^2}{[\eta_T(G-1)+1]^6}\!\int\!\frac{{\rm d}^8\bzeta_S}{\pi ^4}\,
e^{-\bzeta_S^\dagger \bzeta_S/N_S}\left[1 - \frac{|\zeta_{S_{1_H}}+\zeta_{S_{2_H}}|^2}{2N_S}\right]
\left[1 - \frac{|\zeta_{S_{1_V}}-\zeta_{S_{2_V}}|^2}{2N_S}\right]e^{-\bhata_S^\dagger\bzeta_S}e^{\bzeta_S^\dagger\bhata_S}.
\label{chiAstep3}
\end{equation}
\end{widetext}
where $N_S \equiv [\eta_T(G-1)+1]/G$, as given in Sec.~\ref{lossy_BSM}.  Because $[\eta_T(G-1)]^2/[\eta_T(G-1)+1]^6$ is the probability of getting an $I'_{+H}I'_{-V}$ herald, it is immediately obvious that
\begin{widetext}
\begin{equation}
\hat{\rho}_{{\bf S}\mid I'_{+H}I'_{-V}} = \int\!\frac{{\rm d}^8\bzeta_S}{\pi^4}\,
e^{-\bzeta_S^\dagger \bzeta_S/N_S}\left[1 - \frac{|\zeta_{S_{1_H}}+\zeta_{S_{2_H}}|^2}{2N_S}\right]
\left[1 - \frac{|\zeta_{S_{1_V}}-\zeta_{S_{2_V}}|^2}{2N_S}\right]e^{-\bhata_S^\dagger\bzeta_S}e^{\bzeta_S^\dagger\bhata_S},
\label{chiAstep4}
\end{equation}
is the conditional density operator of the ZALM source's output modes given there has been an $I'_{+H}I'_{-V}$ herald.  This result shows that 
\begin{align}
\chi_A^{\rho_{{\bf S}\mid I'_{+H}I'_{-V}}}(\bzeta_S) = e^{-\bzeta_S^\dagger \bzeta_S/N_S}\left[1 - \frac{|\zeta_{S_{1_H}}+\zeta_{S_{2_H}}|^2}{2N_S}\right] \left[1 - \frac{|\zeta_{S_{1_V}}-\zeta_{S_{2_V}}|^2}{2N_S}\right]
\label{chiAstep5}
\end{align}
is the associated anti-normally-ordered characteristic function, as given in Eq.~(\ref{CondxCharFn}).
 
At last we are ready to go after the conditional Bell-state output probabilities of islands-based ZALM with a lossy partial BSM given an $I'_{+H}I'_{-V}$ herald.  These are ${}_{\bf S}\langle \psi^-| \hat{\rho}_{{\bf S}\mid I'_{+H}I'_{-V}} |\psi^-\rangle_{\bf S}$, ${}_{\bf S}\langle \psi^+| \hat{\rho}_{{\bf S}\mid I'_{+H}I'_{-V}} |\psi^+\rangle_{\bf S}$, 
${}_{\bf S}\langle \phi^+| \hat{\rho}_{{\bf S}\mid I'_{+H}I'_{-V}} |\phi^+\rangle_{\bf S}$ and 
${}_{\bf S}\langle \phi^-| \hat{\rho}_{{\bf S}\mid I'_{+H}I'_{-V}} |\phi^-\rangle_{\bf S}$,
where 
\begin{align}
|\psi^\pm\rangle_{\bf S} &\equiv (|1\rangle_{S_{1_H}}|0\rangle_{S_{1_V}}|0\rangle_{S_{2_H}}|1\rangle_{S_{2_V}}  \pm |0\rangle_{S_{1_H}}|1\rangle_{S_{1_V}}|1\rangle_{S_{2_H}}|0\rangle_{S_{2_V}})/\sqrt{2},
\end{align}
and  
\begin{align}
|\phi^\pm\rangle_{\bf S} \equiv (|1\rangle_{S_{1_H}}|0\rangle_{S_{1_V}}|1\rangle_{S_{2_H}}|0\rangle_{S_{2_V}} 
 \pm |0\rangle_{S_{1_H}}|1\rangle_{S_{1_V}}|0\rangle_{S_{2_H}}|1\rangle_{S_{2_V}})/\sqrt{2}.
\end{align}
Employing Eq.~(\ref{chiAstep4}) now results in
\begin{align}
{}_{\bf S}\langle \psi^\pm| \hat{\rho}_{{\bf S}\mid I'_{+H}I'_{-V}} |\psi^\pm\rangle_{\bf S} &=\frac{N_S^4}{2}\int\!{\rm d}^8\bzeta_S\,
\frac{e^{-\bzeta_S^\dagger \bzeta_S/N_S}}{(\pi N_S)^4}\left[1 - \frac{|\zeta_{S_{1_H}}+\zeta_{S_{2_H}}|^2}{2N_S}\right]
\left[1 - \frac{|\zeta_{S_{1_V}}-\zeta_{S_{2_V}}|^2}{2N_S}\right] \nonumber  \\[.05in]
&\times [(1-|\zeta_{S_{1_H}}|^2)(1-|\zeta_{S_{2_V}}|^2) \pm 2{\rm Re}\,(\zeta_{S_{1_H}}\zeta_{S_{1_V}}^*\zeta_{S_{2_H}}^*\zeta_{S_{2_V}}) + 
(1-|\zeta_{S_{1_V}}|^2)(1-|\zeta_{S_{2_H}}|^2)],
\label{successDeriv2}
\end{align}
and 
\begin{align}
{}_{\bf S}\langle \phi^\pm| \hat{\rho}_{{\bf S}\mid I'_{+H}I'_{-V}} |\phi^\pm\rangle_{\bf S} &=\frac{N_S^4}{2}\int\!{\rm d}^8\bzeta_S\,
\frac{e^{-\bzeta_S^\dagger \bzeta_S/N_S}}{(\pi N_S)^4}\left[1 - \frac{|\zeta_{S_{1_H}}+\zeta_{S_{2_H}}|^2}{2N_S}\right]
\left[1 - \frac{|\zeta_{S_{1_V}}-\zeta_{S_{2_V}}|^2}{2N_S}\right] \nonumber  \\[.05in]
&\times [(1-|\zeta_{S_{1_H}}|^2)(1-|\zeta_{S_{2_H}}|^2) \pm 2{\rm Re}\,(\zeta_{S_{1_H}}\zeta_{S_{1_V}}^*\zeta_{S_{2_H}}\zeta_{S_{2_V}}^*) + 
(1-|\zeta_{S_{1_V}}|^2)(1-|\zeta_{S_{2_V}}|^2)]\}.
\label{errorDeriv4}
\end{align}

Multiplying out the polynomial terms Eqs.~(\ref{successDeriv2}), (\ref{errorDeriv4}) and using complex-Gaussian moment factoring~\cite{AppAfootnote}, which applies because $e^{-\bzeta_S^\dagger \bzeta_S/N_S}/(\pi N_S)^4$ is the joint pdf for 4 iid circulo-complex Gaussian random variables with variance $N_S$, we verify Eqs.~(\ref{s_nIV.a}) and (\ref{e_nIV.a}).  To handle the remaining possible herald events we start from Eq.~(\ref{chiAstep2}) and parallel what we did to get $\chi_A^{\rho_{{\bf S}\mid I'_{+H}I'_{-V}}}(\bzeta_S)$ and $\chi_A^{\rho_{{\bf S}\mid I'_{\pm H}I'_{\pm V}}}(\bzeta_S)$.  The results we obtain are
\begin{align}
\chi_A^{\rho_{{\bf S}\mid I'_{+V}I'_{-H}}}(\bzeta_S) = e^{-\bzeta_S^\dagger \bzeta_S/N_S}\left[1 - \frac{|\zeta_{S_{1_H}}-\zeta_{S_{2_H}}|^2}{2N_S}\right]\left[1 - \frac{|\zeta_{S_{1_V}}+\zeta_{S_{2_V}}|^2}{2N_S}\right]
\label{chiAstep6}
\end{align}
and
\begin{align}
\chi_A^{\rho_{{\bf S}\mid I'_{\pm H}I'_{\pm V}}}(\bzeta_S) = e^{-\bzeta_S^\dagger \bzeta_S/N_S}\left[1 - \frac{|\zeta_{S_{1_H}}\pm\zeta_{S_{2_H}}|^2}{2N_S}\right] \left[1 - \frac{|\zeta_{S_{1_V}}\pm\zeta_{S_{2_V}}|^2}{2N_S}\right].
\label{chiAstep7}
\end{align}
\end{widetext}
Evaluating the conditional Bell-state output probabilities for these heralds, again using complex-Gaussian moment factoring, gives us
\begin{equation}
{}_{\bf S}\langle \psi^-| \hat{\rho}_{{\bf S}\mid I'_{-H}I'_{+V}} |\psi^-\rangle_{\bf S} = 
{}_{\bf S}\langle \psi^+| \hat{\rho}_{{\bf S}\mid I'_{\pm H}I'_{\pm V}} |\psi^+ \rangle_{\bf S} = N_S^6/2,
\end{equation}
with all the others equaling zero.  Thus, regardless of what herald occurs the $S_1S_2$ output state from a ZALM QTX with a lossy partial BSM but lossless signal paths has unit Bell-state fidelity.

Turning now to Bell-state fraction for the lossy partial BSM, we already have the Bell-state output probabilities, so we need only find the conditional probability that the $S_1S_2$ output state is loadable, i.e., sends at least one photon to Alice and at least one to Bob.  We will do so for the case of  an $I'_{+H}I'_{-V}$ herald, because the other heralds give the same result, as we found for the output-state probabilities.  

The unnormalized density operator, $\tilde{\rho}_{{\bf S}_\ell\mid I'_{+H}I'_{-V}}$, for the loadable $({\bf S}_\ell$) output states under the preceding conditioning is 
\begin{align}
\tilde{\rho}_{{\bf S}_\ell\mid I'_{+H}I'_{-V}} &= \hat{\rho}_{{\bf S}\mid I'_{+H}I'_{-V}} \nonumber \\[.05in]
&- [{}_{S_1}\langle {\bf 0}|\hat{\rho}_{{\bf S}\mid I'_{+H}I'_{-V}}|{\bf 0}\rangle_{S_1}]\otimes |{\bf 0}\rangle_{S_1}\,{}_{S_1}\langle {\bf 0}| \nonumber \\[.05in]
&- [{}_{S_2}\langle {\bf 0}|\hat{\rho}_{{\bf S}\mid I'_{+H}I'_{-V}}|{\bf 0}\rangle_{S_2}]\otimes |{\bf 0}\rangle_{S_2}\,{}_{S_2}\langle {\bf 0}| \nonumber \\[.05in] &+ [{}_{\bf S}\langle {\bf 0}|\hat{\rho}_{{\bf S}\mid I'_{+H}I'_{-V}}|{\bf 0}\rangle_{\bf S}]|{\bf 0}\rangle_{\bf S}\,{}_{\bf S}\langle {\bf 0}|,
\label{loadableDensOp}
\end{align}   
where $|{\bf 0}\rangle_{S_k}$, for $k = 1,2$, is the vacuum state of the $S_k$ modes and $|{\bf 0}\rangle_{\bf S}$ is the vacuum state of the $S_1S_2$ modes.  We then have that
\begin{equation}
\Pr({\rm loadable}\mid I'_{+H}I'_{-V}) = {\rm Tr}(\tilde{\rho}_{{\bf S}'\mid I'_{+H}I'_{-V}}).
\label{loadableBSM}
\end{equation}
Using Eq.~(\ref{chiAstep4}) and taking the trace in the coherent-state basis results in four integral terms each of which can be evaluated via complex-Gaussian moment factoring.  The result is
\begin{equation}
\Pr({\rm loadable}\mid I'_{+H}I'_{-V}) = 1- N_S^2/2.
\end{equation}

\section{Islands-based ZALM with partial-BSM loss and propagation loss: Derivations \label{AppendB}}
The principal goals of this appendix are to derive the Bell-state fidelity and the Bell-state fraction for islands-based ZALM with a lossy partial-BSM apparatus and lossy propagation.  Fortunately, all the heavy lifting has already been done in Appendix~\ref{AppendA}, so we can proceed quickly.  

When the $S_1S_2$ modes resulting from an $I'_{+H}I'_{-V}$ herald propagate to Alice and Bob's QRXs via transmissivity-$\eta_R$ channels, the $S_AS_B$ modes arriving there have annihilation operators given by
\begin{equation}
\hat{a}_{A_P} = \sqrt{\eta_R}\,\hat{a}_{S_{1_P}} +\sqrt{1-\eta_R}\,\hat{b}_{A_P}, \mbox{ for $P=H,V,$}
\end{equation}
and
\begin{equation}
\hat{a}_{B_P} = \sqrt{\eta_R}\,\hat{a}_{S_{2_P}} +\sqrt{1-\eta_R}\,\hat{b}_{B_P}, \mbox{ for $P=H,V,$}
\end{equation}
where the $\{\hat{b}_{A_P},\hat{b}_{B_P}\}$ modes are all in their vacuum states.  With $\tilde{\bf S}$ denoting the $S_AS_B$ modes, we then have that
\begin{widetext}
\begin{equation}
\chi_A^{\rho_{\tilde{\bf S}\mid I'_{+H}I'_{-V}}}(\bxi) = 
 \chi_A^{\rho_{{\bf S}\mid I'_{+H}I'_{-V}}}(\sqrt{\eta_R}\,\bxi) e^{-(1-\eta_R)\bxi^\dagger\bxi} 
= e^{-\bxi^\dagger \bxi/N'_S}\left[1 - \frac{\eta_R|\xi_{A_H}+\xi_{B_H}|^2}{2N_S}\right] 
 \left[1 - \frac{\eta_R|\xi_{A_V}-\xi_{B_V}|^2}{2N_S}\right],
\label{chiAproploss}
 \end{equation}
where $\bxi^\dagger \equiv \left[\begin{array}{cc} \bxi_A^\dagger & \bxi_B^\dagger\end{array}\right]$,  $\bxi_K^\dagger \equiv \left[\begin{array}{cc}\xi_{K_H}^* & \xi_{K_V}^*\end{array}\right]$ for $K = A,B$, and $N_S' \equiv [\eta_R/N_S + (1-\eta_R)]^{-1}$.  For loss in both the partial BSM and the propagation we now have that
\begin{align}
{}_{\tilde{\bf S}}\langle \psi^\pm|\hat{\rho}_{\tilde{\bf S}\mid I'_{+H}I'_{-V}}|\psi^\pm\rangle_{\tilde{\bf S}} &=\frac{N_S^{\prime 4}}{2}\int\!{\rm d}^8\bxi\,
\frac{e^{-\bxi^\dagger \bxi/N'_S}}{(\pi N'_S)^4}\left[1 - \frac{\eta_R|\xi_{A_H}+\xi_{B_H}|^2}{2N_S}\right]
\left[1 - \frac{\eta_R|\xi_{A_V}-\xi_{B_V}|^2}{2N_S}\right] \nonumber  \\[.05in]
&\times [(1-|\xi_{A_H}|^2)(1-|\xi_{B_V}|^2) \mp 2{\rm Re}\,(\xi_{A_H}\xi_{A_V}^*\xi_{B_H}^*\xi_{B_V}) + 
(1-|\xi_{A_V}|^2)(1-|\xi_{B_H}|^2)],
\end{align}
and
\begin{align}
{}_{\tilde{\bf S}}\langle \phi^\pm|\hat{\rho}_{\tilde{\bf S}\mid I'_{+H}I'_{-V}}|\phi^\pm\rangle_{\tilde{\bf S}} &=\frac{N_S^{\prime 4}}{2}\int\!{\rm d}^8\bxi\,
\frac{e^{-\bxi^\dagger \bxi/N'_S}}{(\pi N'_S)^4}\left[1 - \frac{\eta_R|\xi_{A_H}+\xi_{B_H}|^2}{2N_S}\right]
\left[1 - \frac{\eta_R|\xi_{A_V}-\xi_{B_V}|^2}{2N_S}\right] \nonumber  \\[.05in]
&\times [(1-|\xi_{A_H}|^2)(1-|\xi_{B_H}|^2) \pm 2{\rm Re}\,(\xi_{A_H}\xi_{A_V}^*\xi_{B_H}\xi_{B_V}^*) + 
(1-|\xi_{A_V}|^2)(1-|\xi_{B_V}|^2)].
\end{align}
Multiplying out the polynomial terms and using complex-Gaussian moment factoring we obtain Eqs.~(\ref{PrSnLossProp}) and (\ref{PrEnLossProp}).   

The Bell-state probabilities for the other three heralds are obtained in a like manner from
\begin{align}
{}_{\tilde{\bf S}}\langle \psi^\pm|\hat{\rho}_{\tilde{\bf S}\mid I'_{-H}I'_{+V}}|\psi^\pm\rangle_{\tilde{\bf S}} &=\frac{N_S^{\prime 4}}{2}\int\!{\rm d}^8\bxi\,
\frac{e^{-\bxi^\dagger \bxi/N'_S}}{(\pi N'_S)^4}\left[1 - \frac{\eta_R|\xi_{A_H}-\xi_{B_H}|^2}{2N_S}\right]
\left[1 - \frac{\eta_R|\xi_{A_V}+\xi_{B_V}|^2}{2N_S}\right] \nonumber  \\[.05in]
&\times [(1-|\xi_{A_H}|^2)(1-|\xi_{B_V}|^2) \mp 2{\rm Re}\,(\xi_{A_H}\xi_{A_V}^*\xi_{B_H}^*\xi_{B_V}) + 
(1-|\xi_{A_V}|^2)(1-|\xi_{B_H}|^2)],
\end{align}
\begin{align}
{}_{\tilde{\bf S}}\langle \phi^\pm|\hat{\rho}_{\tilde{\bf S}\mid I'_{-H}I'_{+V}}|\phi^\pm\rangle_{\tilde{\bf S}} &=\frac{N_S^{\prime 4}}{2}\int\!{\rm d}^8\bxi\,
\frac{e^{-\bxi^\dagger \bxi/N'_S}}{(\pi N'_S)^4}\left[1 - \frac{\eta_R|\xi_{A_H}-\xi_{B_H}|^2}{2N_S}\right]
\left[1 - \frac{\eta_R|\xi_{A_V}+\xi_{B_V}|^2}{2N_S}\right] \nonumber  \\[.05in]
&\times [(1-|\xi_{A_H}|^2)(1-|\xi_{B_H}|^2) \pm 2{\rm Re}\,(\xi_{A_H}\xi_{A_V}^*\xi_{B_H}\xi_{B_V}^*) + 
(1-|\xi_{A_V}|^2)(1-|\xi_{B_V}|^2)],
\end{align}
\begin{align}
{}_{\tilde{\bf S}}\langle \psi^\pm|\hat{\rho}_{\tilde{\bf S}\mid I'_{+H}I'_{+V}}|\psi^\pm\rangle_{\tilde{\bf S}} &=\frac{N_S^{\prime 4}}{2}\int\!{\rm d}^8\bxi\,
\frac{e^{-\bxi^\dagger \bxi/N'_S}}{(\pi N'_S)^4}\left[1 - \frac{\eta_R|\xi_{A_H}+\xi_{B_H}|^2}{2N_S}\right]
\left[1 - \frac{\eta_R|\xi_{A_V}+\xi_{B_V}|^2}{2N_S}\right] \nonumber  \\[.05in]
&\times [(1-|\xi_{A_H}|^2)(1-|\xi_{B_V}|^2) \mp 2{\rm Re}\,(\xi_{A_H}\xi_{A_V}^*\xi_{B_H}^*\xi_{B_V}) + 
(1-|\xi_{A_V}|^2)(1-|\xi_{B_H}|^2)],
\end{align}
\begin{align}
{}_{\tilde{\bf S}}\langle \phi^\pm|\hat{\rho}_{\tilde{\bf S}\mid I'_{+H}I'_{+V}}|\phi^\pm\rangle_{\tilde{\bf S}} &=\frac{N_S^{\prime 4}}{2}\int\!{\rm d}^8\bxi\,
\frac{e^{-\bxi^\dagger \bxi/N'_S}}{(\pi N'_S)^4}\left[1 - \frac{\eta_R|\xi_{A_H}+\xi_{B_H}|^2}{2N_S}\right]
\left[1 - \frac{\eta_R|\xi_{A_V}+\xi_{B_V}|^2}{2N_S}\right] \nonumber  \\[.05in]
&\times [(1-|\xi_{A_H}|^2)(1-|\xi_{B_H}|^2) \pm 2{\rm Re}\,(\xi_{A_H}\xi_{A_V}^*\xi_{B_H}\xi_{B_V}^*) + 
(1-|\xi_{A_V}|^2)(1-|\xi_{B_V}|^2)],
\end{align}
\begin{align}
{}_{\tilde{\bf S}}\langle \psi^\pm|\hat{\rho}_{\tilde{\bf S}\mid I'_{-H}I'_{-V}}|\psi^\pm\rangle_{\tilde{\bf S}} &=\frac{N_S^{\prime 4}}{2}\int\!{\rm d}^8\bxi\,
\frac{e^{-\bxi^\dagger \bxi/N'_S}}{(\pi N'_S)^4}\left[1 - \frac{\eta_R|\xi_{A_H}-\xi_{B_H}|^2}{2N_S}\right]
\left[1 - \frac{\eta_R|\xi_{A_V}-\xi_{B_V}|^2}{2N_S}\right] \nonumber  \\[.05in]
&\times [(1-|\xi_{A_H}|^2)(1-|\xi_{B_V}|^2) \mp 2{\rm Re}\,(\xi_{A_H}\xi_{A_V}^*\xi_{B_H}^*\xi_{B_V}) + 
(1-|\xi_{A_V}|^2)(1-|\xi_{B_H}|^2)],
\end{align}
and
\begin{align}
{}_{\tilde{\bf S}}\langle \phi^\pm|\hat{\rho}_{\tilde{\bf S}\mid I'_{-H}I'_{-V}}|\phi^\pm\rangle_{\tilde{\bf S}} &=\frac{N_S^{\prime 4}}{2}\int\!{\rm d}^8\bxi\,
\frac{e^{-\bxi^\dagger \bxi/N'_S}}{(\pi N'_S)^4}\left[1 - \frac{\eta_R|\xi_{A_H}-\xi_{B_H}|^2}{2N_S}\right]
\left[1 - \frac{\eta_R|\xi_{A_V}-\xi_{B_V}|^2}{2N_S}\right] \nonumber  \\[.05in]
&\times [(1-|\xi_{A_H}|^2)(1-|\xi_{B_H}|^2) \pm 2{\rm Re}\,(\xi_{A_H}\xi_{A_V}^*\xi_{B_H}\xi_{B_V}^*) + 
(1-|\xi_{A_V}|^2)(1-|\xi_{B_V}|^2)].
\end{align}
The results gotten in this manner are
\begin{align}
\Pr(|\psi^-\rangle_{\tilde{\bf S}}\mid \psi^-) &= \Pr(|\psi^+\rangle_{\tilde{\bf S}}\mid \psi^+) \nonumber \\[.05in]
&= \frac{N_S'^4}{2}\!\left[2(1-N_S')^2 -\frac{2\eta_R(3N_S'^3-5N_S'^2+2N_S')}{N_S} + \frac{\eta_R^2(5N_S'^4-6N_S'^3+2N_S'^2)}{N_S^2}\right],
\end{align}
and 
\begin{align}
\Pr(|\psi^\pm\rangle_{\tilde{\bf S}}\mid \psi^\mp) &= \Pr(|\phi^\pm\rangle_{\tilde{\bf S}}\mid \psi^\mp) = 
\Pr(|\phi^\pm\rangle_{\tilde{\bf S}}\mid \psi^\pm)\nonumber \\[.05in]
&= \frac{N_S'^4}{2}\!\left[2(1-N_S')^2 -\frac{2\eta_R(3N_S'^3-5N_S'^2+2N_S')}{N_S} + \frac{\eta_R^2(4N_S'^4-6N_S'^3+2N_S'^2)}{N_S^2}\right].
\end{align}

Out next task in this appendix is to find the Bell-state fraction of $\hat{\rho}_{\tilde{\bf S}\mid I'_{+H}I'_{-V}}$.  Given that we already have the Bell-state probabilities in hand, we need only find, cf.~Eqs.~(\ref{loadableDensOp}) and (\ref{loadableBSM}),
\begin{equation}
\Pr({\rm loadable}\mid I'_{+H}I'_{-V}) = {\rm Tr}(\tilde{\rho}_{\tilde{\bf S}_\ell\mid I'_{+H}I'_{-V}}),
\end{equation}
where
\begin{align}
\tilde{\rho}_{\tilde{\bf S}_\ell\mid I'_{+H}I'_{-V}} = &\,\,\hat{\rho}_{\tilde{\bf S}\mid I'_{+H}I'_{-V}} - [{}_{S_A}\langle {\bf 0}|\hat{\rho}_{\tilde{\bf S}\mid I'_{+H}I'_{-V}}|{\bf 0}\rangle_{S_A}]\otimes |{\bf 0}\rangle_{S_A}\,{}_{S_A}\langle {\bf 0}| - [{}_{S_B}\langle {\bf 0}|\hat{\rho}_{\tilde{\bf S}\mid I'_{+H}I'_{-V}}|{\bf 0}\rangle_{S_B}]\otimes |{\bf 0}\rangle_{S_B}\,{}_{S_B}\langle {\bf 0}| \nonumber \\[.05in] & + [{}_{\tilde{\bf S}}\langle {\bf 0}|\hat{\rho}_{\tilde{\bf S}\mid I'_{+H}I'_{-V}}|{\bf 0}\rangle_{\tilde{\bf S}}]|{\bf 0}\rangle_{\tilde{\bf S}}\,{}_{\tilde{\bf S}}\langle {\bf 0}|.
\label{preloadable}
\end{align} 
\end{widetext}
Taking the operator-valued inverse Fourier transform of Eq.~(\ref{chiAproploss}) to obtain $\hat{\rho}_{\tilde{\bf S}\mid I'_{+H}I'_{-V}}$, and then evaluating the Eq.~(\ref{preloadable})'s trace in the coherent-state basis results in four integral terms.  Each of these integrals can be done analytically via complex-Gaussian moment factoring, with their sum being Eq.~(\ref{loadable2}).  The same procedure can be repeated for the other three herald possibilities leading to results given by the right-hand side of Eq.~(\ref{loadable2}).  

Finally, let us sketch how the Bell-state diagonality reported in Sec.~\ref{lossy_prop} can be obtained.  We have that
\begin{widetext}
\begin{align}
{}_{\tilde{\bf S}}\langle \psi^+|\hat{\rho}_{\tilde{\bf S}\mid I'_{+H}I'_{-V}}|\psi^-\rangle_{\tilde{\bf S}} &=\frac{N_S^{\prime 4}}{2}\int\!{\rm d}^8\bxi\,
\frac{e^{-\bxi^\dagger \bxi/N'_S}}{(\pi N'_S)^4}\left[1 - \frac{\eta_R|\xi_{A_H}+\xi_{B_H}|^2}{2N_S}\right]
\left[1 - \frac{\eta_R|\xi_{A_V}-\xi_{B_V}|^2}{2N_S}\right] \nonumber  \\[.05in]
&\times [(1-|\xi_{A_H}|^2)(1-|\xi_{B_V}|^2) - 2i{\rm Im}\,(\xi_{A_H}\xi_{A_V}^*\xi_{B_H}^*\xi_{B_V}) + 
(1-|\xi_{A_V}|^2)(1-|\xi_{B_H}|^2)],
\end{align}
and
\begin{align}
{}_{\tilde{\bf S}}\langle \phi^\pm|\hat{\rho}&_{\tilde{\bf S}\mid I'_{+H}I'_{-V}}|\psi^-\rangle_{\tilde{\bf S}} =\frac{N_S^{\prime 4}}{2}\int\!{\rm d}^8\bxi\,
\frac{e^{-\bxi^\dagger \bxi/N'_S}}{(\pi N'_S)^4}\left[1 - \frac{\eta_R|\xi_{A_H}+\xi_{B_H}|^2}{2N_S}\right]
\left[1 - \frac{\eta_R|\xi_{A_V}-\xi_{B_V}|^2}{2N_S}\right] \nonumber  \\[.05in]
&\times [-(1-|\xi_{A_H}|^2)\xi_{B_H}\xi^*_{B_V} \pm  (1-|\xi_{A_V}|^2)\xi^*_{B_H}\xi_{B_V} 
+\xi_{A_H}\xi^*_{A_V}(1-|\xi_{B_H}|^2) \mp \xi^*_{A_H}\xi_{A_V}(1-|\xi_{B_V}|^2)].
\end{align}
\end{widetext}
Multiplying out the polynomials and doing the complex-Gaussian moment factoring we find that these off-diagonal Bell-state elements of $\hat{\rho}_{\tilde{\bf S}\mid I'_{+H}I'_{-V}}$ vanish.  A similar procedure for the other off-diagonal Bell-state elements of this density operator shows that they vanish too.  Indeed, this procedure 
can be applied to the $\hat{\rho}_{\tilde{\bf S}\mid I'_{-H}I'_{+V}}$ and $\hat{\rho}_{\tilde{\bf S}\mid I'_{\pm H}I'_{\pm V}}$ to show that their off-diagonal Bell-state elements also vanish.  We omit the details.


\begin{thebibliography}{99}
\bibitem{footnote2}Sagnac sources can also be configured to create time-bin entanglement.
\bibitem{footnoteA}Tracing out the signal states from Eq.~(\ref{islands_state}) shows that the $4N_I$ idler modes are in independent, identically-distributed thermal states  whose photon-number distribution is given by  Eq.~(\ref{BoseEinstein}).  Thus, the probability of the dual-Sagnac's $n$th islands emitting a total of 2 idler photon, one of which is $H$ polarized while the other is $V$ polarized, is $4P_0^2P_1^2 = 4(G-1)^2/G^6$, because there are 4 ways to have one $H$ and one $V$ emission with each way occurring with probability $P_0^2P_1^2$.
\bibitem{footnoteB}The probability that the $n$th island does not produce a herald is $1-\Pr(\mathcal{H}_n)$.  Because the $N_I$ islands' emissions are iid, the probability that \emph{none} of the islands produce a herald is $[1-\Pr(\mathcal{H}_n)]^{N_I}$. Thus, the probability that at least one island produces a herald is $1-[1-\Pr(\mathcal{H}_n)]^{N_I}$, and  Eq.~(\ref{true-herald}) then follows because only half of the heralds are true heralds.
\bibitem{footnote4}We will continue to assume the SPDs have no dark counts.
\bibitem{footnoteC}The single-mode thermal state with average photon number $G-1$ is a circulo-complex Gaussian superposition of coherent states, i.e., its density operator is 
\begin{displaymath}
\hat{\rho}_{G-1} = 
= \int\!{\rm d}^2\alpha\,\frac{e^{-|\alpha|^2/(G-1)}}{\pi (G-1)}\,|\alpha\rangle\langle \alpha|,
\end{displaymath}
where the $\{|\alpha\rangle\}$ are coherent states.  
Treating $\eta_T <1$ photodetection as propagation loss through a transmissivity-$\eta_T$ beam splitter followed by unit-efficiency detection, we have that 
\begin{displaymath}
\hat{\rho}_{G-1}^{(\eta_T)} = \int\!{\rm d}^2\alpha\,\frac{e^{-|\alpha|^2/(G-1)}}{\pi (G-1)}\,|\sqrt{\eta_T}\,\alpha\rangle\langle \sqrt{\eta_T}\,\alpha|,
\end{displaymath}
is the density operator for the unit-efficiency detector's illumination, 
because coherent states are just attenuated by propagation loss.  Changing variables to $\beta = \sqrt{\eta_T}\,\alpha$ results in 
\begin{align}
\hat{\rho}_{G-1}^{(\eta_T)} &= \int\!{\rm d}^2\beta\,\frac{e^{-|\beta|^2/\eta_T(G-1)}}{\pi \eta_T(G-1)}\,|\beta\rangle\langle \beta| \nonumber\\[.05in] 
&= \hat{\rho}_{\eta_T(G-1)}, \nonumber
\end{align}
from which Eq.~(\ref{lossy_herald}) follows immediately.
\bibitem{footnote5}Here, and in all that follows, integrals without limits are from $-\infty$ to $\infty$ in all their dimensions.
\bibitem{footnoteD}Because the idler emissions from the ZALM QTX's SPDCs are iid, the probability of an $nm$th cross-island herald is the same as the probability of an $n$th same-island herald.  Hence Eq.~(\ref{nm_herald}) follows directly from Eq.~(\ref{lossy_herald}).
\bibitem{footnote6}Equation~(\ref{true-herald2}) is obtained as follows.  For SPCI heralding 
$\Pr(\mathcal{H}_{\rm true}) = [1-\Pr(\bigcap_{n,m=1}^{N_I}\mathcal{H}^\perp_{nm})]/2$, where $\mathcal{H}_{nm}^\perp$ is the event that there is not an $nm$th island-pair herald.  There will not be an $nm$th island-pair herald when there are no $H$-polarized $n$th-island idler detections,  which occurs with probability $1-\sqrt{\Pr(\mathcal{H}_{nm})}$, or there are no $V$-polarized $m$th-island idler detections, which also occurs with probability $1-\sqrt{\Pr(\mathcal{H}_{nm})}$.  These events, however, are not mutually exclusive.  It  thus follows that $\Pr(\bigcap_{n,m=1}^{N_I}\mathcal{H}^\perp_{nm}) = 2[1-\sqrt{\Pr(\mathcal{H}_{nm})}\,]^{N_I} - [1-\sqrt{\Pr(\mathcal{H}_{nm})}\,]^{2N_I}$, where the negative term corrects for the overcounting of events with neither $n$th-island $H$-polarized nor $m$th-island $V$-polarized idler detections.  The factor of 1/2 in Eq.~(\ref{true-herald2}) excludes false heralds.
\end{thebibliography}
\end{document}